\documentclass[a4paper,11pt]{article}
\pdfoutput=1
\usepackage{jheppub}
\usepackage{amsmath,amssymb,bm,graphicx,epsf,colordvi}
\usepackage{slashed}
\def\bea#1\eea{\begin{align}#1\end{align}} 
\usepackage{diagbox}
\usepackage{mathtools}
\usepackage{lipsum}
\usepackage{soul}
\usepackage{enumitem}
\usepackage{braket}
\usepackage{bm}
\allowdisplaybreaks 
\usepackage{color}
\usepackage[dvipsnames]{xcolor}

\usepackage{mathrsfs}
\usepackage{appendix}
\usepackage{bm}
\usepackage{lipsum}

\newcommand{\bef}{\begin{figure}[hbt]\centering}
\newcommand{\eef}{\end{figure}}

\usepackage{mathtools}
\usepackage{booktabs}
\usepackage{graphicx}
\graphicspath{ {./figure/} }
\usepackage{caption}
\usepackage{lipsum}

\newcommand{\beq}{\begin{equation}}
\newcommand{\eeq}{\end{equation}}
\def\bea#1\eea{\begin{align}#1\end{align}}

\def \be  {\begin{equation}}
\def \ee  {\end{equation}}
\def \ba  {\begin{eqnarray}}
\def \ea  {\end{eqnarray}}

\allowdisplaybreaks

\usepackage{graphicx}

\usepackage{multicol}
\usepackage{subfig}
\usepackage[dvipsnames]{xcolor}
\usepackage{amssymb}
\usepackage{pifont}
\usepackage{multirow}
\usepackage{arydshln}
\usepackage{booktabs}  

\def\Fig#1{Fig.~{\ref{#1}}}
\DeclareRobustCommand{\Sec}[1]{Sec.~\ref{#1}}
\DeclareRobustCommand{\Eq}[1]{Eq.~(\ref{#1})}
\DeclareRobustCommand{\tab}[1]{Table.~\ref{#1}}
\usepackage{cleveref}

\usepackage{stackengine}

\title{Energy-Energy Flow Networks}
\author[a]{Arianna Garcia Caffaro,}
\affiliation[a]{Department of Physics, Yale University, New Haven, CT 06511}

\author[a]{Ian Moult,}

\author[a]{Chase Shimmin}

\emailAdd{arianna.garciacaffaro@yale.edu,ian.moult@yale.edu,chase.shimmin@yale.edu}

\abstract{Jet substructure provides one of the most exciting new approaches for searching for physics in and beyond the Standard Model at the Large Hadron Collider. 
Modern jet substructure searches are often performed with Neural Network (NN) taggers which study the jets’ radiation distributions in great detail, far beyond what is theoretically described by parton shower generators.  
While this represents a great opportunity, as NNs look deeper into the structure of jets they become increasingly sensitive both to perturbative and non-perturbative theoretical uncertainties.
It is therefore important to be able to control which aspects of both regimes the networks focus on, and to develop techniques for quantifying these uncertainties.
In this paper we take two steps in this direction: First, we introduce EnFNs, a generalization of the Energy Flow Networks (EFNs) which directly probes higher point correlations in jets, as motivated by recent advances in the study of energy correlators. Second, we introduce a number of techniques to quantify and visualize their robustness to non-perturbative corrections.
We highlight the importance of such considerations in a toy study incorporating systematics into a search, and maximizing for the network's discovery significance, as opposed to absolute tagging performance.
We hope this study continues the interest in understanding the role QCD systematics play in Machine Learning applications and opens the door to a better interplay between theory and experiment in HEP.  
}

\begin{document}

\maketitle

\section{Introduction}\label{sec:intro}

In recent years, jet substructure has emerged as one of the most exciting developments at the Large Hadron Collider (LHC). While originally proposed to aid the search for $H\to b\bar b$ \cite{Butterworth:2008iy}, it has redefined the scope of jet physics, enabling new precision Standard Model measurements, greatly expanding the reach of Beyond the Standard Model searches, and providing new ways to study nuclear physics. For recent reviews, see \cite{Larkoski:2017jix,Asquith:2018igt,Marzani:2019hun}. Jet substructure itself has been completely transformed by applications of machine learning (ML). As compared to early taggers which used simple high-level jet features, e.g. \cite{Thaler:2011gf,Thaler:2010tr,Larkoski:2014gra,Moult:2016cvt,Larkoski:2014zma,Larkoski:2013eya}, ML based taggers are able to fully exploit the rich internal structure of jets, leading to significantly improved performance \cite{ATLAS:2023nwp,ATLAS:2023zcb,ATLAS:2023krw,Duperrin:2023elp,ATLAS:2022qby,CMS:2020poo}. These new approaches have significantly improved our ability to search for interesting SM phenomena, including
$H\to b\bar b$ \cite{ATLAS:2024doi,ATLAS:2019lwq,ATLAS:2024yzu,ATLAS:2023jdk,ATLAS:2020jwz,CMS:2020zge,ATLAS:2021nsx}, $H\to c \bar c$ \cite{CMS:2022psv}, and di-Higgs production
\cite{CMS:2022gjd,CMS:2024fkb}. ML based approaches have also opened up exciting opportunities with advanced flavor tagging
\cite{Kats:2024eaq,Blekman:2024wyf,Nakai:2020kuu}, and have even enabled new precision QCD measurements with novel approaches to unfolding \cite{Andreassen:2019cjw,H1:2023fzk,ATLAS:2025qtv,Song:2023sxb}. For reviews, see \cite{Schwartz:2021ftp,Karagiorgi:2022qnh}. While these new ML based approaches represent a tremendous opportunity for improving the power of searches, they must be used with caution, since they exploit features of the internal structure of jets at a level far beyond what is theoretically understood. It is therefore important to understand the uncertainties associated with these ML based approaches, how they manifest in searches, and how networks can be guided towards features whose theoretical description can be systematically improved.

Jets have a complicated structure, which involves both perturbative and non-perturbative components.  In the last decade, there has been tremendous progress in improving our understanding of jets. Advancements have been primarily made in understanding the perturbative structure of jets, these come in the form of improvements in parton shower generators \cite{Dasgupta:2020fwr,Hoche:2024dee}, as well as in analytic calculations. Excitingly, there has been progress towards the incorporation of higher-point splitting functions \cite{Catani:1998nv,Campbell:1997hg} into parton shower generators \cite{Hoche:2017iem}, and analytic calculations of observables sensitive to higher point correlations within jets \cite{Chen:2019bpb,Komiske:2022enw,Chen:2022swd,Chicherin:2024ifn}. These offer a starting point from which to improve the description of multi-point correlations within jets. However, it is important that these theoretical improvements translate to advancements in experimentally used search strategies. For ML approaches this is non-trivial, since although our description of the perturbative substructure of jets has improved tremendously, our description of the non-perturbative physics of jets, in particular of hadronization, remains quite poor.  ML approaches, without further guidance, can dangerously focus on the non-perturbative aspects of jets. This is highly problematic, as it not only implies the tremendous advances in theory are not having an impact in experimental measurements, but also indicates that searches performed using such networks carry large uncertainties. We would therefore like to highlight the direct connection between ML architectures and improvements in our perturbative description of jets, as well as develop ways to both control and quantify the sensitivity to non-perturbative physics. Several interesting recent studies which focus on the interplay between ``robustness" to non-perturbative effects and performance of the network include \cite{Proceedings:2018jsb,Butter:2022xyj,ATLAS:2024rua}.

In this paper we take several steps in these directions. First, we introduce a new class of network architectures for studies of jet substructure, which directly targets higher point correlations in jets, motivated by recent progress in the study of energy correlator observables \cite{Moult:2025nhu}. These networks are based on the Energy Flow Networks (EFNs) and Particle Flow Networks (PFNs) \cite{Komiske:2018cqr}, but generalizes them to consider higher point combinations (pairs, triplets, ...) of particles. Using a generalization of the EFN architecture, we are able to make them infrared and collinear (IRC) safe by construction. We name these the EnFN and PnFN, combining the nomenclature for E$^{\text{n}}$C for n-point energy correlator observables \cite{Chen:2020vvp} with that for energy flow networks. We study the performance of these networks, compared to the well tested ParticleNet \cite{Qu:2019gqs} benchmark. While the EnFN networks are slightly less performant due to their restriction to IRC safe information, the PnFN networks perform competitively. An interesting feature of these networks is their interpretability since they are directly looking at correlation lengths at a given scale.

The second major focus of this paper is on introducing techniques to quantify and characterize the resilience of networks to non-perturbative effects. Hence, we develop an approach which uses the same Pythia events, before and after hadronization, to quantify the network's sensitivity to the hadronization process of individual events. We show how Pareto plots \cite{10.1007/978-3-540-88908-3_14} serve as a convenient tool for visualizing the robustness vs. performance of ML taggers. The use of Pareto plots to study the resilience of ML based jet taggers has also recently appeared in \cite{Gambhir:2025xim}. To introduce an additional handle for studying the impact of non-perturbative corrections, we consider the network's response to jets which have first been clustered into sub-jets of small radii.  Aside from introducing an additional layer of perturbative regularization, this methodology also improves the computational time required for higher point networks, similar to the case of energy correlators \cite{Budhraja:2024xiq}.

Finally, we emphasize how uncertainties due to non-perturbative modeling can impact the choice of networks when optimizing for search significance. In a simple Z search, we show that optimizing based on search significance can drastically modify the effective performance of ML based taggers.  We believe that the techniques that we have introduced will be useful more generally in studies of jet substructure tagging, and we hope that they motivate more detailed studies of ML systematics in searches.

An outline of this paper is as follows. In \Sec{sec:ir_safety} we review the EFN networks, and their infrared and collinear safety; we then introduce the  generalized EnFN and PnFN networks which directly incorporate higher point correlations. Additionally, we discuss different possible inputs for the networks, as well as their perturbative regularization through clustering.
In \Sec{sec:network_studies} we study the performance of the E2FN and P2FN networks on different inputs, and comment on their interpretability.
In \Sec{sec:roboustness} we study the performance and robustness of an array of networks. First, we compare their raw performance to the well established benchmark, ParticleNet \cite{Qu:2019gqs}. Next, we study the robustness of these networks to hadronization corrections, introducing Pareto plots as a convenient way for visualizing the robustness vs. performance trade-off. We also perform a simple toy study to highlight the importance of properly incorporating systematic uncertainties to maximize significance.  Concluding remarks are made in \Sec{sec:conclusions}

\section{Generalized Networks with N-Point Correlations}\label{sec:ir_safety}

In this section we introduce a new family of networks, which we term  PnFN and EnFN, combining the nomenclature for E$^N$C for N-point energy correlator observables \cite{Chen:2020vvp} with that for energy flow and particle flow networks (EFNs and PFNs) \cite{Komiske:2018cqr}. The goal of these networks is to focus on multi-particle correlations within jets in an infrared and collinear safe manner, motivated by recent theoretical progress in the study of energy correlators \cite{Moult:2025nhu}.

Infrared and collinear (IRC) safety is a property which ensures a jet’s observable or, in this case, a network's output will remain unchanged under soft emissions and collinear splittings.  This property is a necessary condition for calculating an observable in perturbative QCD, as it guarantees that all the real and virtual corrections of an interaction cancel to give a finite result.  In general, networks used for jet tagging are not IRC safe. A key step was achieved in \cite{Komiske:2018cqr}, by introducing the Energy Flow Network (EFN), which explicitly built IRC safety into the structure of the network. We will review its explicit implementation shortly. In the spirit of energy correlators, the EFN can be thought of as a form of one-point correlator, or heat map of the energy flux in a jet. However, much of the information about a jet is directly encoded in its higher point correlation functions. We are therefore motivated to generalize the EFN networks to directly target higher point correlations. A nice feature of multi-point correlations is that they naturally introduce an associated scale, corresponding to the angular scale of the correlation. Since many physical problems in jet tagging have associated angular scales, structuring a network around such correlations may help it focus its attention on physical features relevant at these natural scales. Furthermore, we believe these networks offer a promising avenue for establishing connections with analytic studies of multi-point correlation functions in jets. With these motivations in mind, we are led to introduce and explore our new EnFN and PnFN networks.

Before introducing these new networks, we begin by briefly reviewing the structure of Energy Flow and Particle Flow Networks. Energy Flow Networks and Particle Flow Networks treat particles as an unordered, variable-length set.  These general architectures employ a per-particle function, $\Phi$, and a latent space function $F$ to learn key features from the event. The main differences between the two architectures stem from the fact that the EFN was designed to be IRC safe.  To that end, the EFN takes all energy dependence outside of the $\Phi$ function,  making it depend solely on the particles' angular information, $\hat\eta_i$.  This change forces the NN to scale linearly with energy and ensures that low energy particles contribute less to the sum, thus making the network IRC safe.  This structure change is apparent when comparing \Eq{eq:PFN} with \Eq{eq:EFN}, where $e_i$ is the energy of the $i$th particle, $\hat\eta_i$ is its angular information, and $p_i$ is any feature attributed to it.  In these equations, both $\Phi$ and $F$ denote neural networks, where $\Phi$ processes the individual particle information, and $F$ aggregates the outputs of $\Phi$ across all particles, typically via a summation. Explicitly, the PFN takes the form
\begin{align}
\label{eq:PFN}
  \text{PFN} = F \left( \sum_{i}^\text{N} \Phi\left( {p}_i \right) \right)\,,
\end{align}
while the EFN takes the form
\begin{align}
\label{eq:EFN}
  \text{EFN} = F \left( \sum_i^\text{N} e_i \Phi\left( \hat{n}_i \right) \right)\,.
\end{align}
While IRC safety is a desirable feature, the constraints which ensure the EFN’s IRC safety also hamper its performance.  In an attempt to recover some of that performance while maintaining the IRC safety properties, \Sec{ParticleCorrelations} will propose a new NN architecture which extends the EFN to include particle correlations. 

\subsection{New Particle Correlating Architectures}
\label{ParticleCorrelations}

It is well established that a network's performance is largely correlated to the quantity and quality of the data provided to it.  Hence it is no surprise that most state-of-the-art NNs rely on their access to pairwise information.  High energy physics is no different. Given that networks in this field are often trained on distributions which arise from particle interactions, relevant pairwise information is inherently encoded in the form of particle correlations. Such particle correlations contain meaningful and often necessary information to fully understand the intricate nature of any given interaction.  This is particularly true for jets, for which the angular correlations between particles shed light on the perturbative nature of a particle's splitting. Following this notion, in the coming sections we will extend the definition of both the EFN and PFN to include particle correlations. 

\subsubsection{Energy-Energy Flow Network\label{subsec:e2fn}}

In order to improve the EFN's performance while maintaining its IRC safety properties, we extend the notion of EFNs to a particle correlating EnFN, which allows for correlations between $n$ particles.  The EnFN architecture is based on Tkachov’s work of C-continous observables \cite{Tkachov:1999py,Tkachov:1995kk,Sveshnikov:1995vi}. Tkachov defines an energy correlator of $n$ particles as 
\begin{equation}
\label{ECs}
    F_n(P) = \sum_{a_1...a_n} E_{a_1}...E_{a_n}{f_n}(\hat{p}_{a_1},...,\hat{p}_{a_n})\,,
\end{equation}
where $E_{a_n}$ and $\hat{p}_{a_n}$ are the energy and angular information of the $n$-th particle; and ${f}_n$ is a symmetric and continuous function of $n$ arguments \cite{Tkachov:1995kk}.  
Here, $n=1$ corresponds to an energy correlator which is analogous in form to the EFN.  Similarly, an energy correlator of $n$ particles is analogous in form to the EnFN we now propose.  Thus, the EnFN’s structure is given by
\begin{equation}
\label{EnFN}
    \mathrm{EnFN} = \mathrm{F}\left(  \sum_{ij...n}^\mathrm{N} e_i e_j...e_n\Phi(\hat{p}_i,\hat{p}_j,...,\hat{p}_n)\right)\,,
\end{equation}
where $\hat{p}_i$ and $e_i$ are the angular information and energy of the $i$-th particle, respectively; and $\Phi$ is a neural network which ought to be a sufficiently smooth function of its $n$ inputs, in order to maintain C-continuity.

\begin{figure}
\begin{center}
\includegraphics[keepaspectratio, width=\linewidth]{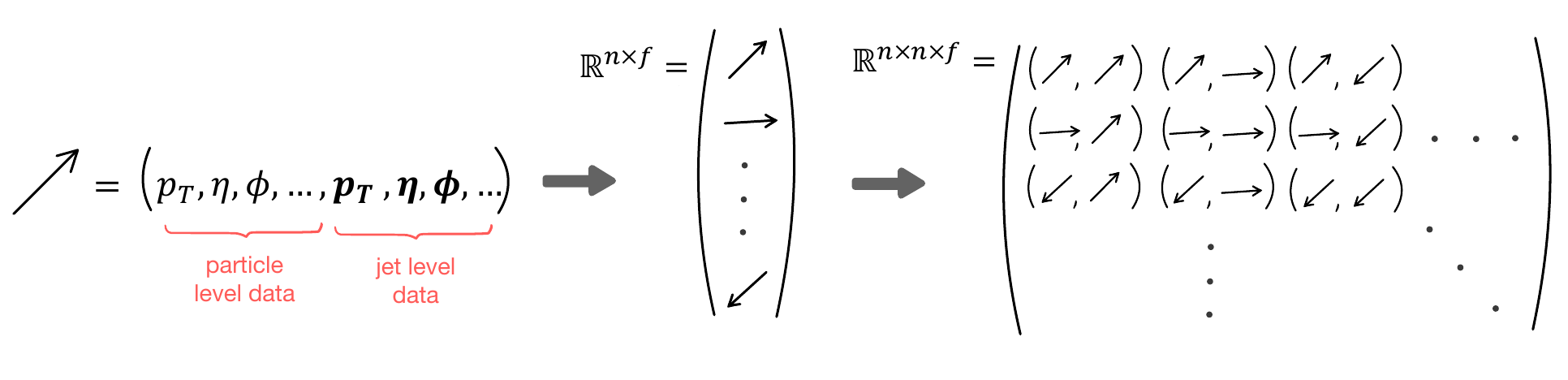} 
\end{center}
\caption{A graphical representation of the inputs fed to the $\Phi$ network, and their pairwise structure.  The features can include any IRC safe information, such as a jet constituent's angular information, as well as jet-level data.  These features live in a space $\mathbb{R}^{n\times f}$ which is then combined in a pairwise manner with features from other particles within the same jet; thus creating a  $\mathbb{R}^{n\times n \times f}$ input structure.}
\label{fig:NN-data-prep}
\end{figure}

\begin{figure}
\begin{center}
\includegraphics[keepaspectratio, width=\linewidth]{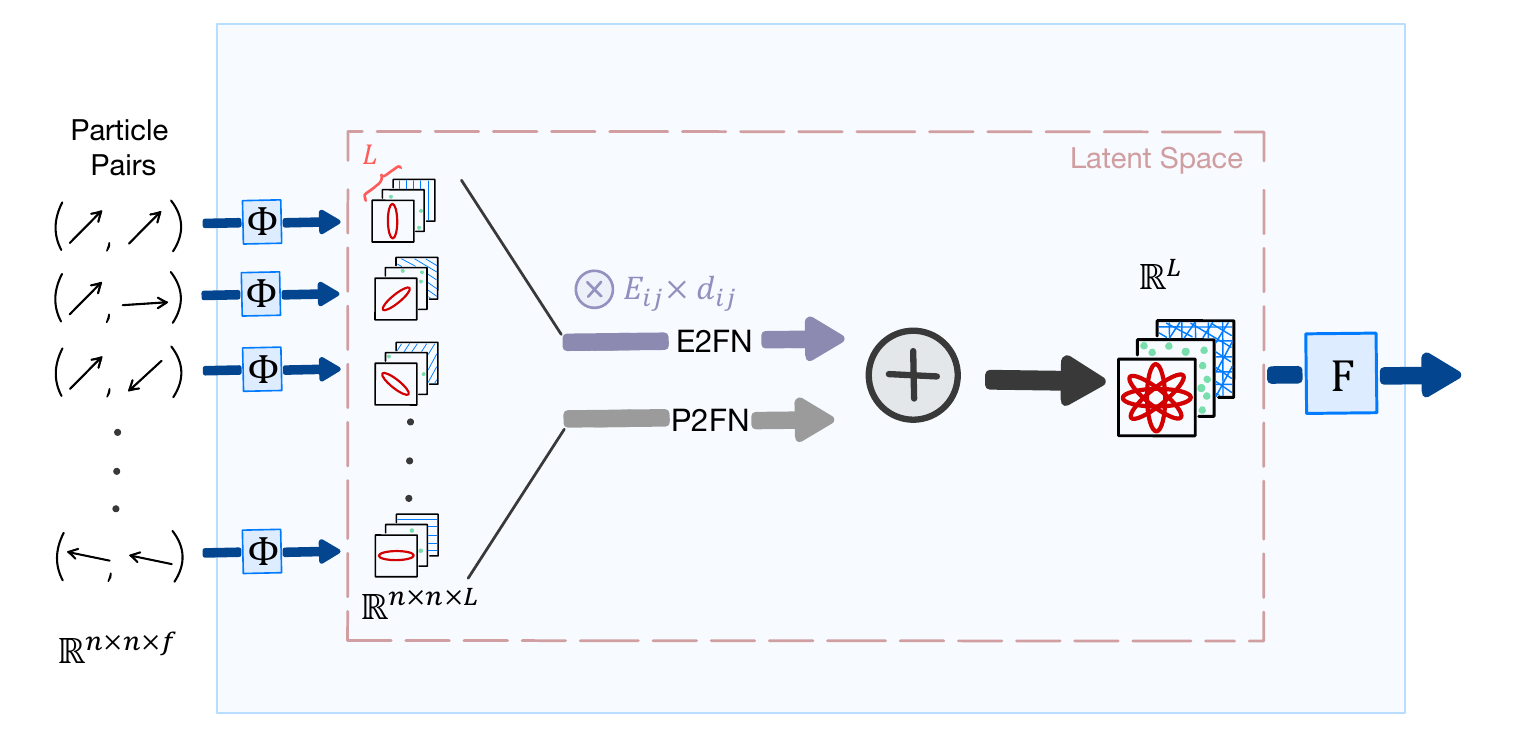} 
\end{center}
\caption{A graphical representation of the data flow logic for an E2FN and P2FN network.  The input for both networks is given by a set of features defined for any pair of particles.  This input is then passed through a $\Phi $ network which maps them into latent space, $\Phi:\mathbb{R}^{n \times n \times f} \rightarrow \mathbb{R}^{n \times n \times L}$.  In the case of the E2FN, the outputs of the $\Phi$ network are then multiplied by the corresponding pairwise energy and damping factor, in order to ensure IRC safety.  On the other hand, the P2FN requires no such modifications and its $\Phi$ outputs can be directly summed over all particle pairs.  Finally, the sum is given as input to the $F$ network which maps the data from the latent space back to the desired output.  For illustration purposes, this is chosen as $F:\mathbb{R}^L \rightarrow \mathbb{R}^2$. } 
\label{fig:NN-data-flow}
\end{figure} 

For the purposes of this paper we will focus on the two particle correlator, E2FN, given by
\begin{equation}
\label{E2FN}
    \mathrm{E2FN} = \mathrm{F}\left(  \sum_{ij}^\mathrm{N} e_i e_j \Phi(\hat{p}_i,\hat{p}_j)\right)\,.
\end{equation}
This network structure inherits the EFN's IRC safety properties.  To demonstrate this, let's look at the case where the two particles being studied undergo a soft emission such that 
\begin{equation}
\begin{aligned}
\label{e2fn-ir}
 \mathrm{E2FN} = \mathrm{F} \left(e_i e_j\Phi\left( \hat{n}_i,\hat{n}_j \right) \right) \; \; \xrightarrow{\text{IR emission}} \; \;
     &\mathrm{F} (e_{i_a} e_{j_a}\Phi\left( \hat{n}_{i_a} \hat{n}_{j_a} \right) + e_{i_b} e_{j_b}\Phi\left( \hat{n}_{i_a} \hat{n}_{j_b} \right) \\ 
     &+ e_{i_b} e_{j_a}\Phi\left( \hat{n}_{i_b} \hat{n}_{j_a} \right) + e_{i_a} e_{j_b}\Phi\left( \hat{n}_{i_a} \hat{n}_{j_b} \right) 
     \\
     &+ e_{i_a} e_{i_b}\Phi\left( \hat{n}_{i_a} \hat{n}_{i_b} \right) + e_{j_a} e_{j_b}\Phi\left( \hat{n}_{j_a} \hat{n}_{j_b} \right) )
\end{aligned}
\end{equation}
Being a soft emission, $e_{i_a}, e_{j_a} \gg e_{i_b}, e_{j_b}$. This allows us to write the relationship between the particle’s energies as:
$e_i =  e_{i_a} + e_{i_b}  \approx  e_{i_a} $. Furthermore, a soft emission implies that the angular direction change of the harder particles is negligible, such that $\hat{n}_i, \hat{n}_j \approx  \hat{n}_{i_a}, \hat{n}_{j_a}$. Therefore, in the limit of a soft emission (i.e. as $ e_{i_b}, e_{j_b} \rightarrow 0$), the last five terms in the sum \Eq{e2fn-ir}  can be dropped and the original relationship is recovered; thus showing the E2FN is infrared safe.
\begin{equation}
\begin{aligned}
     \lim_{e_{i_b}, e_{j_b}\to 0} & \mathrm{F} (e_{i_a} e_{j_a}\Phi\left( \hat{n}_{i_a} \hat{n}_{j_a} \right) + e_{i_b} e_{j_b}\Phi\left( \hat{n}_{i_a} \hat{n}_{j_b} \right)  + e_{i_b} e_{j_a}\Phi\left( \hat{n}_{i_a} \hat{n}_{j_a} \right) \\ & + e_{i_b} e_{j_a}\Phi\left( \hat{n}_{i_a} \hat{n}_{j_b} \right) + e_{i_a} e_{i_b}\Phi\left( \hat{n}_{i_a} \hat{n}_{i_b} \right) + e_{j_a} e_{j_b}\Phi\left( \hat{n}_{j_a} \hat{n}_{j_b} \right) ) \\
     &= \mathrm{F} (e_{i_a} e_{j_a}\Phi\left( \hat{n}_{i_a} \hat{n}_{j_a} \right)) 
    \; 
    \approx
    \; 
    \mathrm{F} (e_{i} e_{j}\Phi\left( \hat{n}_{i} \hat{n}_{j} \right)) 
\end{aligned}
\end{equation}

Similarly, allowing our particle to undergo a collinear splitting as such:
\begin{equation}
\begin{aligned}
\label{e2fn-c}
 \mathrm{E2FN} = \mathrm{F} \left(e_i e_j\Phi\left( \hat{n}_i,\hat{n}_j \right) \right) \; \; \xrightarrow{\text{collinear splitting}} \; \;
     &\mathrm{F} (e_{i_a} e_{j_a}\Phi\left( \hat{n}_{i_a} \hat{n}_{j_a} \right) + e_{i_b} e_{j_b}\Phi\left( \hat{n}_{i_a} \hat{n}_{j_b} \right) \\ 
     &+ e_{i_b} e_{j_a}\Phi\left( \hat{n}_{i_b} \hat{n}_{j_a} \right) + e_{i_a} e_{j_b}\Phi\left( \hat{n}_{i_a} \hat{n}_{j_b} \right) 
     \\
     & + e_{i_a} e_{i_b}\Phi\left( \hat{n}_{i_a} \hat{n}_{i_b} \right) + e_{j_a} e_{j_b}\Phi\left( \hat{n}_{j_a} \hat{n}_{j_b} \right) )
\end{aligned}
\end{equation}
The collinearity of the particles implies $\hat{n}_{i_a} \approx \hat{n}_{i_b} \approx \hat{n}_i$.  Furthermore, by energy conservation, $e_{i_a}+ e_{i_b} = e_i$.  Using both of these relationships and the equivalent relationships for $e_j$, \Eq{e2fn-c} can be re-written as 

\begin{equation}
\begin{aligned}
    &\mathrm{F} (e_{i_a} e_{j_a}\Phi\left( \hat{n}_{i} \hat{n}_{j} \right) + e_{i_b} e_{j_b}\Phi\left( \hat{n}_{i} \hat{n}_{j} \right)  + e_{i_b} e_{j_a}\Phi\left( \hat{n}_{i} \hat{n}_{j} \right) 
    \\
    & + e_{i_a} e_{j_b}\Phi\left( \hat{n}_{i} \hat{n}_{j} \right) + e_{i_a} e_{i_b}\Phi\left( \hat{n}_{i} \hat{n}_{i} \right) + e_{j_a} e_{j_b}\Phi\left( \hat{n}_{j} \hat{n}_{j} \right) )
    \\
    & = \mathrm{F} ((e_{i_a} + e_{i_b}) (e_{j_a}+e_{j_b})\Phi\left( \hat{n}_{i} \hat{n}_{j} \right) + e_{i_a} e_{i_b}\Phi\left( \hat{n}_{i} \hat{n}_{i} \right) + e_{j_a} e_{j_b}\Phi\left( \hat{n}_{j} \hat{n}_{j} \right) )
    \\
    &
    = \mathrm{F} ( e_{i} e_{j}\Phi\left( \hat{n}_{i} \hat{n}_{j} \right) + e_{i_a} e_{i_b}\Phi\left( \hat{n}_{i} \hat{n}_{i} \right) + e_{j_a} e_{j_b}\Phi\left( \hat{n}_{j} \hat{n}_{j} \right) )
\end{aligned}
\end{equation}
 Requiring $\Phi \rightarrow 0$ as the studied particles become collinear, the original equation is recovered.  Hence, showing the E2FN is IRC safe as long as two conditions are met: $\Phi$ must be a continuous function, and $\Phi \rightarrow 0$ as the two studied particles become collinear.

 Using Tkachov's work as a baseline, one can then turn to the more practical details of building a NN following the specifications laid out above.  The E2FN architecture is strongly informed by that of the EFN \Eq{eq:EFN}.  In fact, most architecture features will remain unchanged with two exceptions: the $\Phi$ function transitions into a per-particle-pair function, and its output is damped in order to insure IRC safety.  
 
 Let us then take a look at the full network structure and setup, from input selection to output generation.  Following the equations above, the inputs are given by the particles' angular information.  However, the $\Phi$ network can also take any other IRC safe features as input, such as those corresponding to jet-level information.  Once an array of particle-level and jet-level features has been chosen, they can be combined in a pairwise manner and arranged in matrix form to be fed into $\Phi$, see \Fig{fig:NN-data-prep}. The $\Phi$ network then encodes this pairwise information with feature dimension $f$ into a latent space $L$, such that $\Phi: \mathbb{R}^{n \times n \times f} \rightarrow \mathbb{R}^{n \times n \times L}$.  Following \Eq{E2FN}, the $\Phi$ network's output is then multiplied by the corresponding energy term.  Subsequently, before proceeding to sum over all particle pairs, one must ensure that $\Phi \rightarrow 0$ as the particle pairs become collinear.  This is achieved by damping $\Phi$'s output by the sigmoid function
 \begin{equation}
    \label{eq:damping}
    d\; (k_T) = \frac{\left(\frac{k_T}{\tau}\right)^{\frac{1}{w}}}{\left(\frac{k_T}{\tau}\right)^{\frac{1}{w}} + 1}\,,
\end{equation}
where $w$ is the damping coefficient which parametrizes how fast $d$ drops to zero, and $\tau$ is the damping mean which parametrizes the value of $k_T$ at which $d$ has dropped by half.   For the test studies in coming sections, $w$ and $\tau$ were set to 0.3 and 2.7 GeV, respectively.
Once the damping factor is applied one can proceed to sum the $\Phi$ output over all particle pairs, and feed that sum into a final $F$ network which maps the latent space into the desired output (i.e. $F:\mathbb{R}^L \rightarrow \mathbb{R}^2$) See \Fig{fig:NN-data-flow}.

\subsubsection{Particle-Particle Flow Network}

In addition to generalizing the EFN network to the EnFN, it is also straightforward to generalize the PFN network to incorporate multi-particle correlations, which we term the PnFN. The information flow for this network is very similar to that of the E2FN, discussed in the previous subsection, with the one difference that the P2FN's $\Phi$ outputs need not be multiplied by the energy or damped in anyway, see \Fig{fig:NN-data-flow}.  Since the P2FN is not generally an IRC safe architecture, it has less restrictions on both the inputs it can take and the treatment of the data once inside the network, allowing it higher learning flexibility.

An $n$ correlating PFN takes the general form
\begin{equation}
    \mathrm{PnFN} = \mathrm{F}\left(  \sum_{ij...n}^\mathrm{N} \Phi({p}_i,{p}_j,...,p_n)\right)\,,
\end{equation}
where $\Phi$ can now depend on any particle information, $p_i$, without restrictions.  
For this paper we will focus on the two-particle correlating P2FN, whose explicit form is
\begin{equation}
    \mathrm{P2FN} = \mathrm{F}\left(  \sum_{ij}^\mathrm{N} \Phi({p}_i,{p}_j)\right)\,.
\end{equation}
 We will compare the performance of the P2FN and E2FN networks in \Sec{sec:network_studies}.

\subsection{Perturbative Regularization Through Clustering}\label{sec:regularization}

While the EnFN and PnFN networks are conceptually nice in that their architectures focus on multi-particle correlations, they also become significantly more computationally intensive to train as N increases. For this reason, we will primarily focus on the case of the E2FN and P2FN networks in this paper. However, we also highlight how this computational bottleneck can be overcome by re-clustering the jet constituents. In addition to decreasing the computational burden, re-clustering is a proven method to perturbatively regularize networks, and will play an important role in our studies of robustness to non-perturbative effects, by introducing an additional handle. We therefore discuss this re-clustering approach in more detail.

The complexity of computing multi-point correlations in high-multiplicity jets has also arisen in the study of multi-point energy correlators. In this case, an efficient approach to overcome this issue is to first re-cluster the jet into smaller subjets, and then compute the resulting correlator on these subjets \cite{Budhraja:2024xiq}. In the regime where the angular correlations are much larger than the subjet radius, this has minimal effect. However, at small angular scales, a regime particularly sensitive to non-perturbative corrections, the subjet scale introduces an additional regularization parameter. This is particularly convenient in the case of NNs, since if implemented correctly, it can absorb ill-modeled features inherent of hadronization, and in turn make the network more robust to non-perturbative uncertainties. 

For these studies, jet re-clustering was performed by momentum splitting using an exclusive $k_{T}$ algorithm.  The motivation behind applying this particular re-clustering method arises from the understanding  that there exists a visible transition between the perturbative and non-perturbative regimes which occurs at a specific angular scale. Such angular scale, $\Delta\Theta$, can be related back to the momentum splitting of the constituents, $k_{T}$ if multiplied by the $p_T$ of the jet \Eq{kt}
\begin{equation}
\label{kt}
    k_{T} = p_{T_{jet}}\Delta\Theta\,.
\end{equation}
Several different values of $k_{T}$ in the ranges of $[0.3, 2.5]$ GeV were tested.  A detailed study of the network's performance, and its robustness to hadronization effects for different values of $k_T$ can be found in \Sec{sec:rms}. For the tests in sections \Sec{sec:network_studies}, and \Sec{sec:performance} a momentum splitting cutoff of  1.3 GeV was chosen.

\section{N-Point Network Studies}\label{sec:network_studies}

In this section, we will evaluate the performance of our newly introduced E2FN and P2FN networks, studying their dependence on the nature of the kinematic inputs, as well as the interpretability of their filters.  
After understanding the behavior of the E2FN and P2FN networks, in \Sec{sec:performance} we will compare them with ParticleNet \cite{Qu:2019gqs}, which constitutes a well established benchmark in the field.

\subsection{Dataset Generation}\label{sec:mc}

As a simple physics task to study the performance of our networks, we consider the case of discriminating boosted hadronically decaying Z bosons from a QCD background. Data was generated using the \textsc{Pythia8} \cite{Sjostrand:2007gs,Sjostrand:2014zea} parton shower. We produced events for both the the $Z+$ jets process, with the Z decaying hadronically, which we took as signal, and QCD jet production, which we took as background. Jets were clustered with the anti-$k_T$ algorithm \cite{Cacciari:2008gp} using \textsc{FASTJET} \cite{Cacciari:2011ma}. We considered $R=1$ jets with a $p_T$ cut of $300$ GeV to ensure that the hadronically decaying Z bosons were sufficiently boosted.   Prior to training the NNs, the data was pre-processed.  The angular coordinates of the jet's constituents were centered to the jet's $\eta$ and $\phi$.  Similarly, the constituent's $p_T$ were normalized to the jet's $p_T$.  As discussed in \Sec{sec:regularization}, for certain network configurations a pre-clustering step was performed to apply a momentum splitting cutoff. Our networks were trained with \textsc{Pytorch} \cite{pytorch:2019neurips}.

In order to study both the performance and the robustness of NNs against non-perturbative systematics, each generated event is saved twice: once at parton-level, at the end of the parton shower, and once at hadron-level, after the hadronization is finished.  This allows us to study at an event-by-event level, the impact of hadronization, and will be key to our later studies of the robustness of our architectures.

\subsection{Performance Dependence on Inputs}

Among the networks discussed in \Sec{sec:ir_safety}, those with particle-correlating architectures present a more compelling case for studying the performance input dependence, as they allow for a richer combination of potential inputs.  Hence, this section will focus solely on the study of the E2FN and P2FN's performance. 

The aforementioned networks were tested with variations on the six main inputs listed on Table \ref{table:input_options}.  It is worth highlighting that all but one of the constituent-level inputs --- namely, the constituent's $p_T$ --- are IRC safe, and thus can be given to the E2FN's $\Phi$ network without issue \Eq{E2FN}.  Similarly, it should be noted that global jet variables, such as the $p_{T_{jet}}$ are inherently IRC safe quantities, and hence can also be included as input to the E2FN.  On the other hand, the P2FN has no IRC restrictions on its $\Phi$ network, and can therefore take the $p_T$ of each constituent as an extra input. Taking all this into account, eight $\Phi$ input variations were tested as given in Table \ref{table:input_variations}. 

\begin{figure}
    \centering
    \subfloat[E2FN]{
        \includegraphics[width=.45\linewidth, trim=6mm 0mm 14mm 14mm, clip]{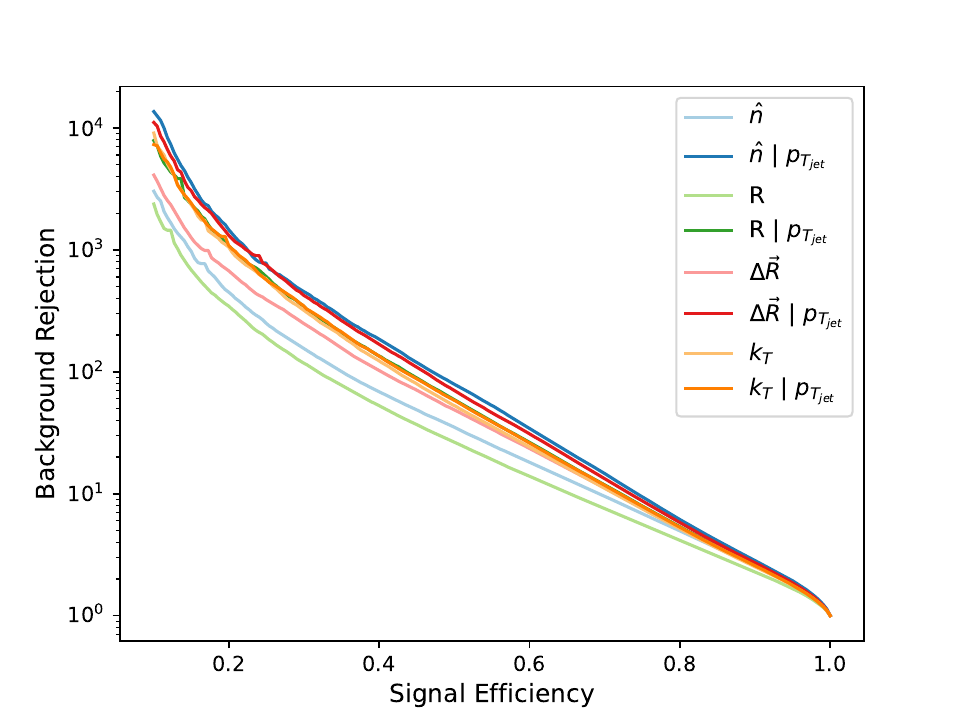}
        \label{subfig:sig-a}
    }\quad
    \subfloat[P2FN]{
        \includegraphics[width=.45\linewidth, trim=6mm 0mm 14mm 14mm, clip]{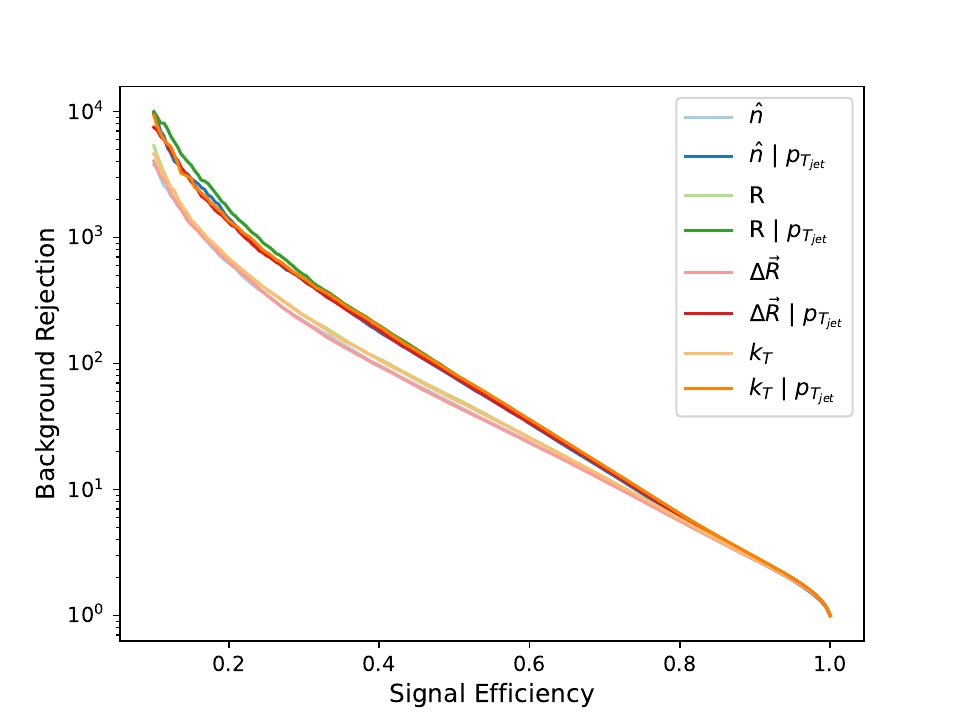}
        \label{subfig:sig-b}
    }
    \caption{The performance of the E2FN and P2FN networks given eight $\Phi$ input variations.
    The signal efficiency versus background rejection curves illustrate how the performance of the particle-correlating architectures vary with different $\Phi$ inputs.  As is to be expected, both architectures exhibit improved performance when $p_{T_{jet}}$ is included as an additional input, likely due to the useful kinematic context such a variable provides. }
    \label{fig:roc_input_var}
\end{figure}

The ROC curves in \Fig{fig:roc_input_var} show an overall competitive performance for the E2FN when compared to the P2FN network.  The input dependence for both networks was studied in two cases: without explicit global jet variables, and with global jet variables.  For the first case, both networks performed best when given $k_T$ as an input.  This is an expected outcome, as $k_T$ contains implicit global jet information which constitutes a considerable advantage over other tested input combinations.  Moreover, the choice for this kinematic variable is well motivated, as it is a good approximation for the constituent’s angular momentum splitting and
hence provides important information for better resolving jet substructures relevant to the network’s task.

However, once global variables are explicitly added as inputs, the performance dependence differs for the two tested networks.  In particular, the P2FN performs best when given high-level variables, such as $\Delta R$ and $k_T$, while the E2FN performs best when given lower level variables such as the vector distance, or the complete angular information.  This trend could be attributed to the fact that the P2FN has a more flexible architecture which allows it to fully exploit the information contained within the high-level variables, while the E2FN, which has a more constrained architecture, benefits from the freedom of manipulating the low level variables as needed to achieve the best results.

For the tests shown on \Sec{sec:roboustness}, the best performing E2FN and P2FN input variations were selected.  These correspond to an $\hat{\eta} \;|\; p_{T_{jet}}$ input for the E2FN; a $\Delta {R} \;|\; p_{T_{jet}}$ input for the P2FN; and  the full angular inputs for the non-particle correlating architectures PFN and EFN.

\begin{table}
\small
\caption{Baseline inputs considered for the E2FN and P2FN.}
  \centering
   \resizebox{0.7\textwidth}{!}{
  \begin{tabular}{ccc}
    \toprule
    \toprule
    Considered Variables & Notation & Definition   \\
    \midrule
    \midrule
    \multicolumn{3}{c}{Constituent-level inputs}   \\
    \hdashline
    \vspace{-2ex}
    \\
    Angular information &$\hat{n}_i,\; \hat{n}_j $  &  $\left( \eta_i,\; \eta_j,\; \phi_i, \;\phi_j\right)$   
    \\
    \vspace{-3ex}
    \\
    Scalar distance & $\Delta R$  &  $ \sqrt{\left( \eta_i - \eta_j\right)^2 + \left( \phi_i - \phi_j\right)^2}$ 
   \\
    \vspace{-3ex}
    \\
     Vector distance & $\Delta \Vec{R}$  &  $\left(\; \left( \eta_i - \eta_j\right),\; \left( \phi_i - \phi_j\right)\;\right)$ 
   \\
    \vspace{-3ex}
    \\
    Momentum splitting & $k_T$  &  $\Delta \Vec{R}*{p_T}_{jet}$ 
   \\
    \vspace{-3ex}
    \\
     Constituent transverse momentum & $p_{T_i}, \; p_{T_j}$  &  $\left(p_{T_i}, \; p_{T_j}\right)$ 
   \\
    \vspace{-2ex}
    \\
    \hdashline
    \multicolumn{3}{c}{Jet-level inputs}   \\
    \hdashline
    \vspace{-2ex}
    \\
     Jet transverse momentum & $p_{T_{jet}}$  &  $\left(p_{T_{jet}}\right)$ 
    \\
  \bottomrule
  \end{tabular}}
  \label{table:input_options}
 \end{table}

\begin{table}
\small
 \caption{$\Phi$ input variations studied}
  \centering
   \resizebox{0.8\textwidth}{!}{
  \begin{tabular}{cccc}
    \toprule
    Naming Scheme & Notation & E2FN inputs & P2FN inputs   \\
    \midrule
    Angular &$\hat{n}$  &  $\left( \hat{n}_i,\; \hat{n}_j\right)$   
    & $\left( \hat{n}_i,\; \hat{n}_j, \;p_{T_i}, \;p_{T_j}\right)$   
    \\
    Conditional Angular &$\hat{n}\;|\;p_{T_{jet}}$  &  $\left( \hat{n}_i,\; \hat{n}_j, \;p_{T_{jet}}\right)$   
    & $\left( \hat{n}_i,\; \hat{n}_j, \;p_{T_i}, \;p_{T_j}, \;p_{T_{jet}}\right)$   
    \\
    \vspace{-2.4ex}
    \\
    Scalar distance & $\Delta R$   &  $\left(\Delta R \right)$   &  $\left( \Delta R,  \;p_{T_i}, \;p_{T_j}\right)$   
    \\
    Conditional Scalar distance & $\Delta R|\;p_{T_{jet}}$ &  $\left(\Delta R, \;p_{T_{jet}}\right)$   & $\left( \Delta R, \;p_{T_i}, \;p_{T_j}, \;p_{T_{jet}}\right)$ 
    \\
    \vspace{-2.4ex}
    \\
    Vector distance & $\Delta \Vec{R}$   &  $\left(\Delta \Vec{R} \right)$   &  $\left( \Delta \Vec{R},  \;p_{T_i}, \;p_{T_j}\right)$   
    \\
    Conditional Vector distance & $\Delta\Vec{R}|\;p_{T_{jet}}$ &  $\left(\Delta \Vec{R}, \;p_{T_{jet}}\right)$   & $\left( \Delta \Vec{R}, \;p_{T_i}, \;p_{T_j}, \;p_{T_{jet}}\right)$ 
    \\
    \vspace{-2.4ex}
    \\
    Momentum splitting & $k_T$   &  $\left(k_T \right)$   &  $\left( k_T,  \;p_{T_i}, \;p_{T_j}\right)$   
    \\
    Conditional Momentum splitting & $k_T|\;p_{T_{jet}}$   &  $\left(k_T, \;p_{T_{jet}}\right)$   &  $\left( k_T,  \;p_{T_i}, \;p_{T_j}, \;p_{T_{jet}}\right)$ 
    \\
   
  \bottomrule
  \end{tabular}}
  \label{table:input_variations}
\end{table}

\begin{table}
\small
 \caption{E2FN and P2FN performance as a function of input variable combinations.}
  \centering
  \resizebox{0.9\textwidth}{!}{
  \begin{tabular}{c|cccc|cccc}
   & \multicolumn{4}{c|}{E2FN} & \multicolumn{4}{c}{P2FN} \\
   \toprule
    \toprule
      & \multirow{2}{0.8cm}{ACC} & \multirow{2}{0.8cm}{AUC} & \multicolumn{2}{c|}{$1/\varepsilon_b$} & \multirow{2}{0.8cm}{ACC} & \multirow{2}{0.8cm}{AUC} & \multicolumn{2}{c}{$1/\varepsilon_b$} \\
      \cmidrule{4-5}  \cmidrule{8-9}& & &$\varepsilon_s=50\%$ & $\varepsilon_s=30\%$  & & &$\varepsilon_s=50\%$ & $\varepsilon_s=30\%$  \\
   \toprule
     $\hat{n}$ & 0.9124 & 0.9742 & 339 & 1438 & 0.9171 & 0.9775 & 459 & 1994\\
     $\Delta R$ & 0.9009 & 0.9672 & 229 & 917 & 0.9177 & 0.9777 & 546 & 2392\\
     $\Delta \Vec{R}$ & 0.9130 & 0.9754 & 560 & 2451 & 0.9167 & 0.9771 & 460 & 1759\\
     \boldsymbol{$k_T$} & \textbf{0.9057} & \textbf{0.9712} & \textbf{689} & \textbf{3560} & \textbf{0.9182} & \textbf{0.9779} & \textbf{573} & \textbf{2355}\\
     \toprule
      \boldsymbol{$\hat{n}\;|\;p_{T_{jet}}$} & \textbf{0.9201} & \textbf{0.9792} & \textbf{1236} & \textbf{6501} & 0.9203 & 0.9796 & 1083 & 4672\\
     \boldsymbol{$\Delta R\;|\;p_{T_{jet}}$} & 0.9107 & 0.9740 & 885 & 4531 & \textbf{0.9203} & \textbf{0.9797} & \textbf{1306} & \textbf{8306}\\
     $\Delta \Vec{R}\;|\;p_{T_{jet}}$ & 0.9172 & 0.9778  & 1159 & 5068 & 0.9203 & 0.9795 & 1080 & 4671\\
     $k_T\;|\;p_{T_{jet}}$ & 0.9099 & 0.9737 & 880 & 4984 & 0.9214 & 0.99802 & 1323 & 7668
     \\
     
\bottomrule\bottomrule
  \end{tabular}}
  \label{table:input_var_perf}
\end{table}
\normalsize

\subsection{Interpretability}\label{sec:interp}

In order to better understand the inner workings of the E2FN, in this section we will take a deeper look at the network's filters.  In particular, we will consider the $\Phi$ network's 1D filters when given $\Delta R$ as an input, and the 2D filters when given $\Delta \Vec{R}$ as input. 

While it is in general difficult to interpret the filters of a network, one potentially interesting feature of building a network based on correlations, such as the EnFN, is that we expect the network to directly identify the correlation length (angular scale) associated with features of interest, and focus the filters' attention onto this angular scale. This approach to the design of the network could be useful in its training. In our particular case, this is scale is extremely simple, namely it is the angular scale, $\theta \sim M_Z/p_T$ of the decaying Z-boson. However, more generally we think that this might be an interesting approach, and might highlight additional interesting scales, such as the scale of deadcone radiation for massive particles, or any of the plethora of scales in nuclear collisions. We hope to explore this further in future work.

The architecture of the E2FN presented in this paper is such that the latent space $L$ of the features, right before they are passed to the $F$ network \Fig{fig:NN-data-flow}, is set to 32.  Hence, these 32 filters of a trained network can be studied to identify which features might be guiding the NN's decision.  

\begin{figure}
    \centering
        \includegraphics[width=0.8\linewidth]{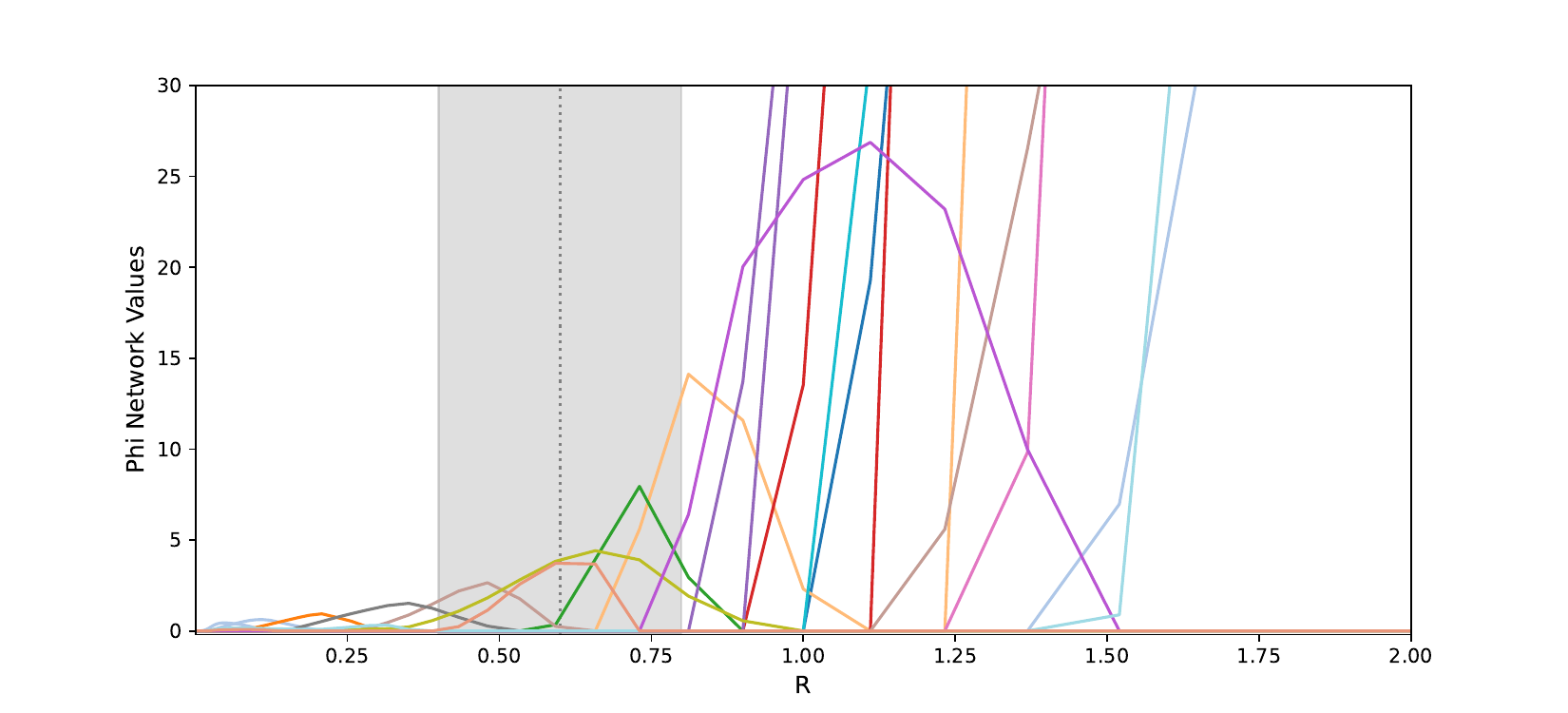}
        \label{subfig:R}
    \caption{An illustration of 32 of the E2FN's  $\Phi$ network filters, when trained with the scalar distance $R$ on boosted Z boson events. The vertical dashed line denotes the angular scale of the boosted Z boson decay. Multiple filters are active in this region, highlighting that the network correctly identifies this characteristic angular scale.  }
    \label{fig:1d_scan}
\end{figure}

Starting with the 1D filters, in \Fig{fig:1d_scan} we show the 32 filters when the network was trained using $\Delta R$ as input. We have also illustrated with the vertical gray line the angular scale associated with the Z-boson decay, which is the characteristic angular scale of the problem. We expect the network to focus on this angular scale, and therefore to have multiple filters active in this region. We see that this is indeed the case, providing us confidence that our network is correctly finding the relevant scale of the problem.

\begin{figure}
\begin{center}
\includegraphics[width=0.8\linewidth]{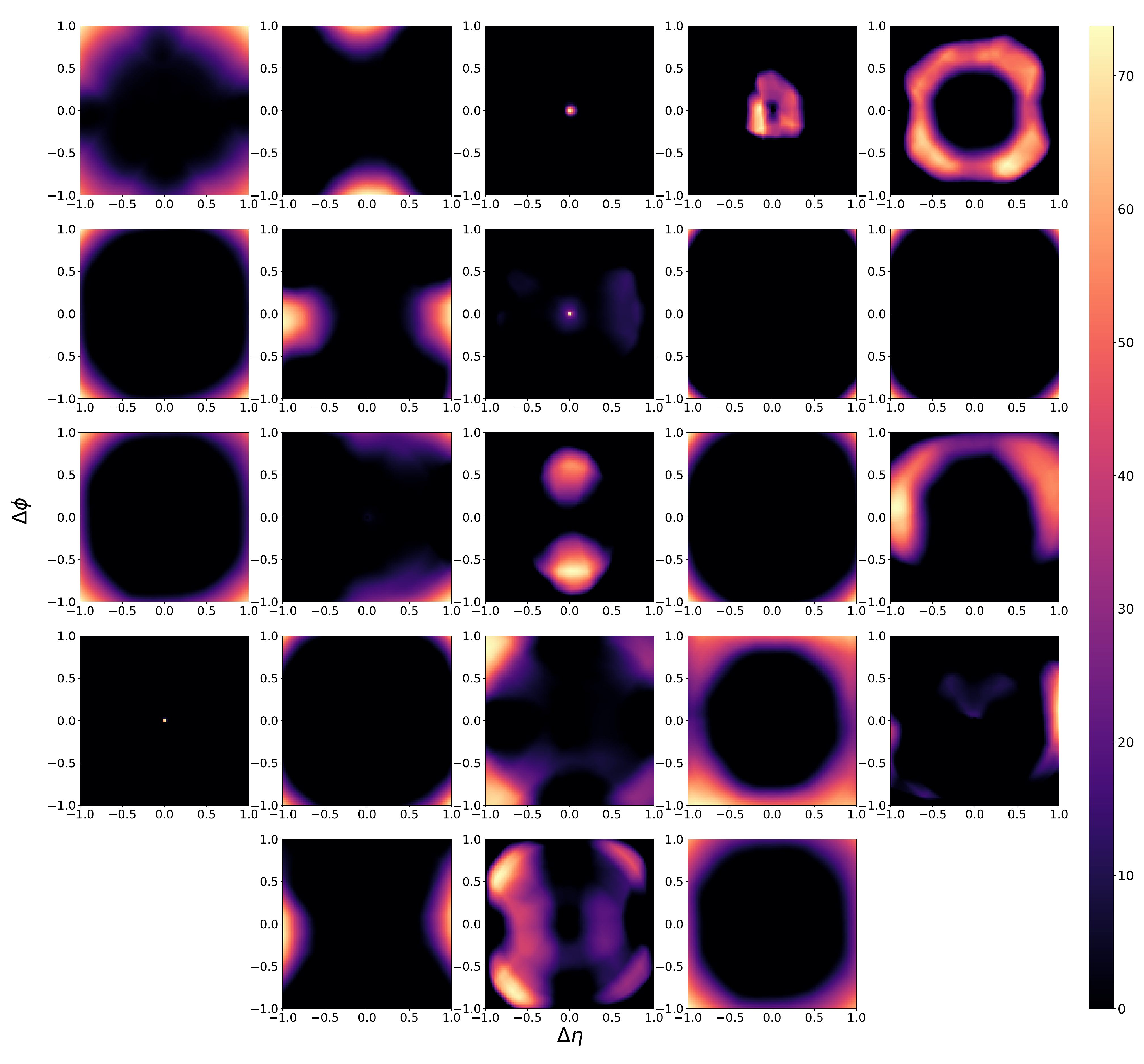}  
\end{center}
\caption{Scan of the E2FN's $\Phi$ Network, when trained with the vector distance as input.  The network is damped according to \Eq{eq:damping} with $\tau$ and $\omega$ set to 0.03 and 0.25, respectively.  For simplicity, only the non-zero $\Phi$ layers are shown in the figure.}
\label{fig:2d_scan}
\end{figure}

For the network trained on $\Delta \Vec{R}$, one can now look at a two-dimensional filter space which scans over $\Delta \eta$ and $\Delta \phi$. Several such filters are shown in \Fig{fig:2d_scan}. The most recurrent features include filters with a sole active cluster at or near the center of the jet's angular coordinates, as well as networks with different angular shapes at the characteristic angular scale of the Z-decay. These suggest, as expected, that the network is placing significant attention on characterizing the shape of radiation at the characteristic angular scale of the Z-decay.

It would be interesting to explore if these patterns could be understood from an analytic perspective, through the calculation of multi-point correlators. Some progress towards analytically understanding the case of quark-gluon discrimination was achieved in
\cite{Kasieczka:2020nyd}.

\section{Performance and Comparison with Other Architectures}\label{sec:performance}

\begin{figure}
\begin{center}
\includegraphics[width=0.8\linewidth]{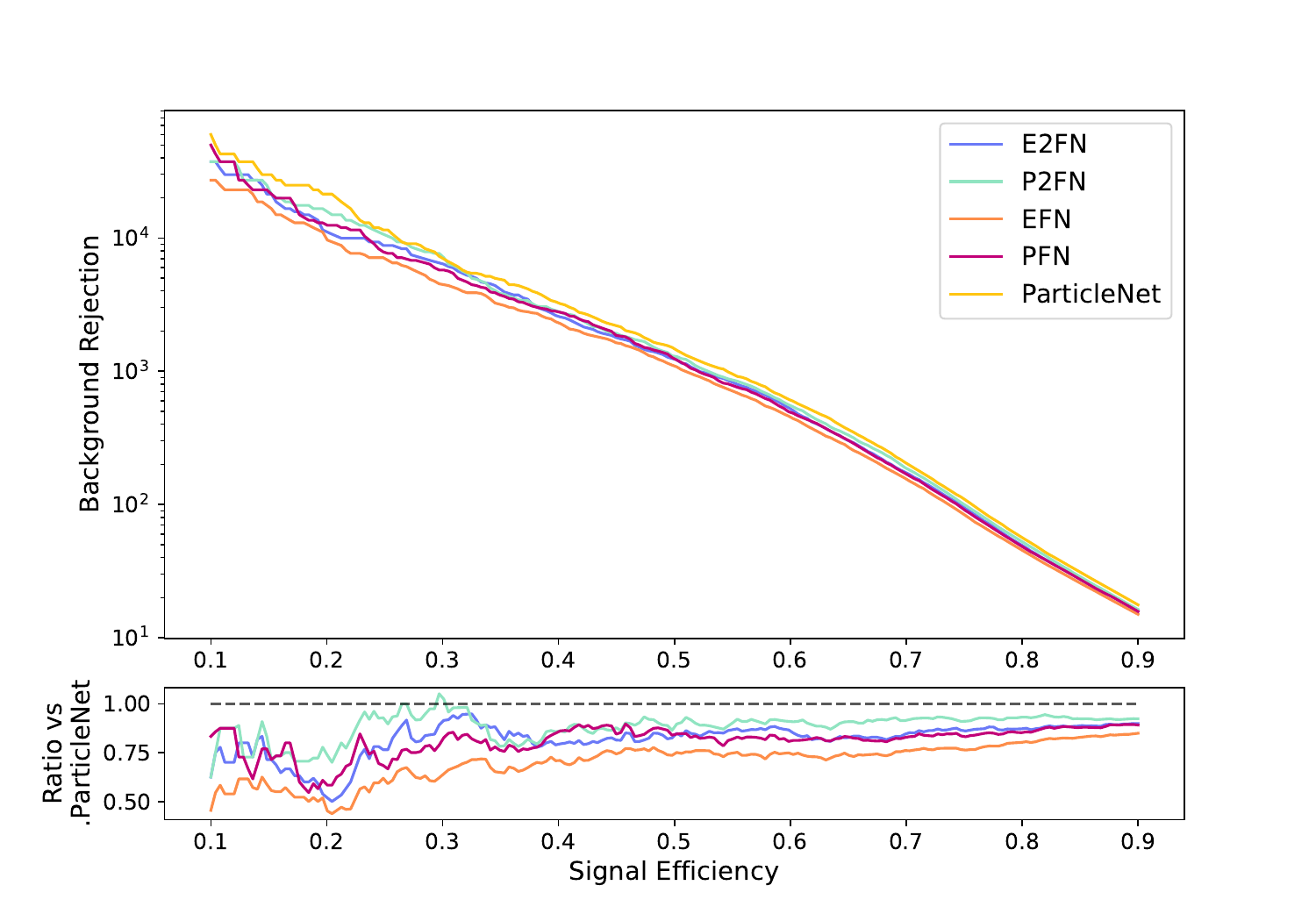} 
\end{center}
\caption{Raw performance comparison of E2FN, EFN, P2FN, PFN against ParticleNet for boosted Z boson tagging in QCD background. The E2FN and P2FN architectures, which directly build in higher point correlations, outperform their EFN and PFN counterparts.}
\label{fig:performance}
\end{figure}

In this section we compare the performance of our newly introduced E2FN and P2FN networks against the well established benchmarks of the EFN and PFN \cite{Komiske:2018cqr}, and ParticleNet\cite{Qu:2019gqs}. ParticleNet was chosen as a baseline as it constitutes one of the most commonly used NNs in the field. 

For the following tests the E2FN and P2FN were given the inputs found to be optimal in \Sec{sec:network_studies}, namely the E2FN was given as input the angular information as well as the transverse momentum of the jet, while the P2FN was given the scalar angular distance between the jet constituents, as well as the transverse momentum of the jet, and the constituent's $p_T$.  The EFN and PFN received the analogous information for a single jet constituent.  ParticleNet was in turn given the angular information and transverse momentum of each particle as well as the $p_T$ of the jet. With these inputs, it constructs a graph using the jet constituents' angular coordinates, where each particle node contains the full particle features as well as jet transverse momentum, see \tab{tab:inputs}.

The performance of the EFN, E2FN, PFN, and P2FN at tagging boosted Z bosons in a QCD background is shown in \Fig{fig:performance}.  Their performances are compared to that of ParticleNet.  As expected, we find that the networks which directly incorporate higher point correlations, namely the P2FN and E2FN, beat their EFN and PFN counterparts. Additionally, as expected, the non-IRC safe ParticleNet and P2FN, which have less restrictive architectures, are the best performing.  These are followed by the particle-correlating E2FN and the non-particle correlating PFN.  Last is the EFN, which is not able to exploit particle correlations nor flexible enough to outperform any other architecture.  While this is an expected outcome, it serves to highlight how including particle correlations can allow IRC safe networks to regain some of the performance lost by nature of their constrained structure. A key message of this paper, however, is that one should not focus entirely on raw performance. Thus, the next sections discuss how the different networks' performances are affected by non-perturbative corrections, and offer methods to quantify their robustness against such theoretical uncertainties.

\begin{table}
\label{test_inputs}
\small
 \caption{$\Phi$ inputs for studied architectures}
  \centering
  \begin{tabular}{cc}
    \midrule
    Architecture   &  $\Phi$ Inputs   \\
    \midrule
     EFN  &  $\left\{\; \hat{\eta}_i,\; {p_T}_{jet}\;\right\}$
    \\
    E2FN  &  $\left\{\; \hat{\eta}_i,\; \hat{\eta}_j,\; {p_T}_{jet}\;\right\}$
    \\
    PFN  &  $\left\{\; \hat{\eta}_i,\; {p_T}_{i}, \;{p_T}_{jet}\;\right\}$
    \\
    P2FN  &  $\left\{\; \Delta R,\; {p_T}_{i}, \;{p_T}_j,\;{p_T}_{jet}\;\right\}$
    \\
     \multirow{2}{*}{ParticleNet} 
       & Node Features: $\left\{ \; \hat{\eta}_i,\;  {p_T}_{i},\; {p_T}_{jet} \right\}$\\
       & Graph Structure: $k$-nearest neighbors in $\hat{\eta}_i$ space
    \\
  \bottomrule
  \end{tabular}
  \label{tab:inputs}
\end{table}
\normalsize

\section{Performance and Robustness of Energy-Energy Flow Networks}\label{sec:roboustness}

While utilizing an IRC safe NN is considered good practice, IRC safety by itself is not enough.  In fact, the non-perturbative corrections of an IRC safe object can still be arbitrarily large.  Hence, in order to assess a measurement's susceptibility to non-perturbative systematics, it is imperative to be able to quantify the robustness of the NN against non-perturbative effects, such as hadronization.  This section will outline two methods for quantifying a NN's robustness. In \Sec{sec:rms} we will propose a new metric for quantifying the robustness of a network to non-perturbative effects, and show that it can be effectively visualized using Pareto plots, while \Sec{sec:toys} will analyze the impact these non-perturbative  uncertainties could have in an analysis.  For a detailed discussion on the data used in these studies please refer to \Sec{sec:mc}.

Several recent studies have focused on the interplay between robustness and resilience
\cite{Proceedings:2018jsb,Butter:2022xyj,ATLAS:2024rua,Gambhir:2025xim}. We hope that this paper, along with these studies, further motivates future studies aimed at understanding and controlling the type of information the NNs are sensitive to, and ensuring the reliability of substructure studies.

\subsection{Quantifying Robustness}\label{sec:rms}

In order to study the robustness of the NNs against non-perturbative effects, each NN is trained on hadron-level data and tested on parton-level data.  This allows for a direct measurement of the change in a NN's output when non-perturbative information is present in its input.  We thus propose a metric for robustness given by the Root Mean Square (RMS) of the difference between a NN's prediction given hadron-level data and parton-level data. Explicitly,
\begin{equation}
\label{eq:rms}
    \mathrm{RMS} = \sqrt{ \frac{\sum\limits_{i}^\mathrm{N} {\Bigl( \;P_i(x\;|\;\text{hadron-data})-P_i(x\;|\;\text{parton-data})\;\Bigl)}^{2}}{\mathrm{N}}}\,.
\end{equation}
The expectation is that a NN which is less sensitive to non-perturbative effects will see less difference between the hadron and parton level data and, hence, will achieve a lower RMS.

\begin{figure}
    \centering
    \includegraphics[keepaspectratio, width=\linewidth, clip]{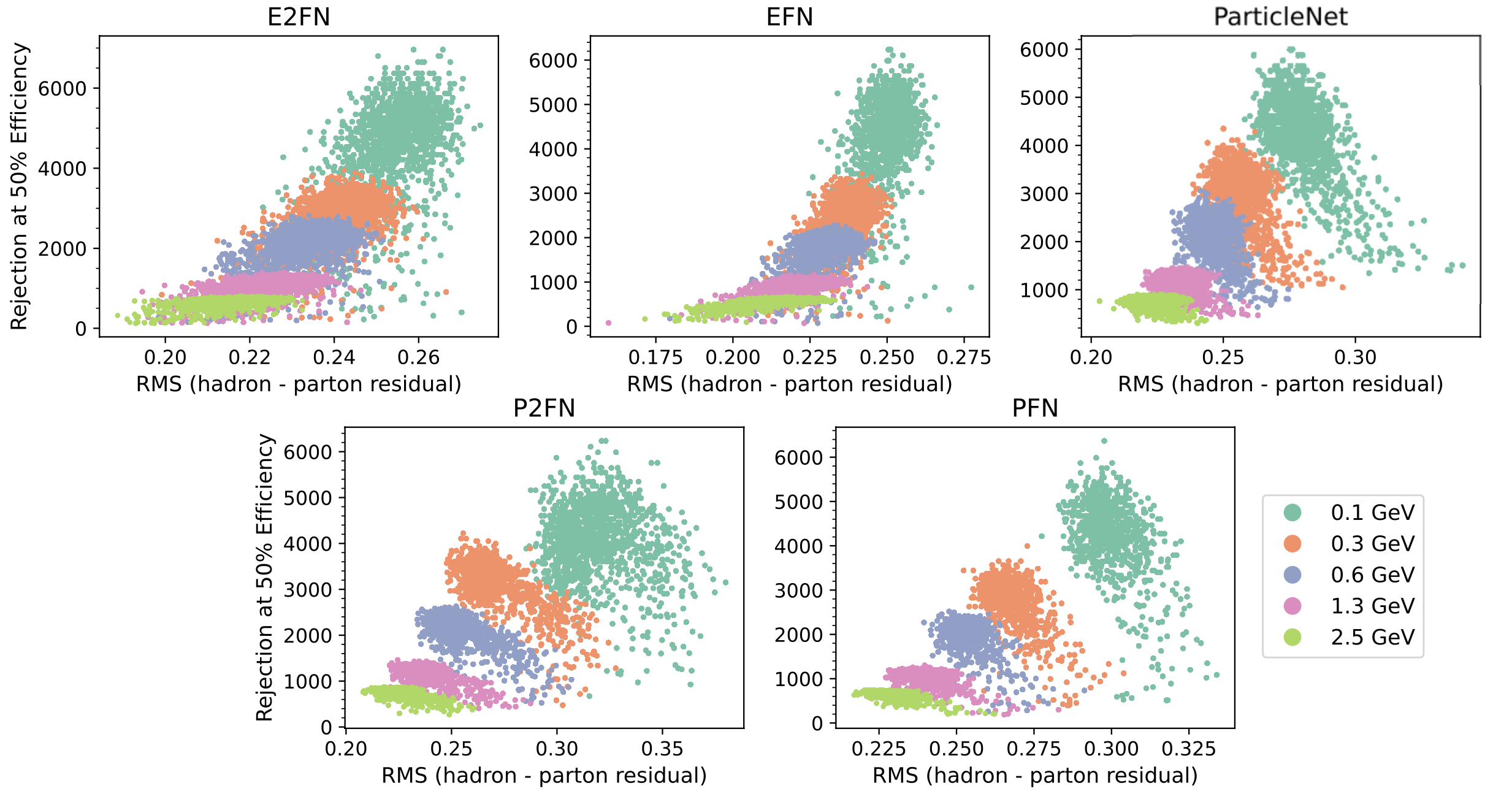}
    \caption{Study of momentum splitting reclustering and its effects on the performance and robustness of NNs}
    \label{fig:recluster}
\end{figure}

The first study performed using this metric focused on analyzing the robustness dependence on clustering radius.  To this end, five networks were trained on five datasets re-clustered using the exclusive $k_{T}$ algorithm with five different momentum splitting cutoffs: 0.1, 0.3, 0.6, 1.3, and 2.5 GeV.  As seen in \Fig{fig:recluster}, the extent to which the jet constituents were re-clustered resulted in a trade-off between robustness and performance.  This is easily understood, as a highly re-clustered output tends to cancel out the non-perturbative details while also to erasing potentially valuable information for jet tagging.  Moreover, this figure highlights that the IRC-safe EFN and E2FN exhibit the highest robustness among all evaluated networks, even when trained to their optimal performance levels.  Another interesting aspect of this study, however, becomes apparent on the trends the different architectures present as they increase their training.  Note that, while the E2FN and EFN tend to increase their robustness as their performances decrease, ParticleNet, PFN and P2FN do not.  For these networks the robustness is either maintained or in some cases increased as the training improves, however even their best robustness remains small in comparison to those of the IRC-safe networks. Overall the EFN and E2FN provide a constrained robustness spectrum shifted towards high robustness, even as the performance increases.  This feature would enable the user to choose a more robust working point if their study required it.  

\begin{figure}[!htb]
   \begin{minipage}{0.48\textwidth}
     \centering
     \includegraphics[height= 5cm, keepaspectratio,]{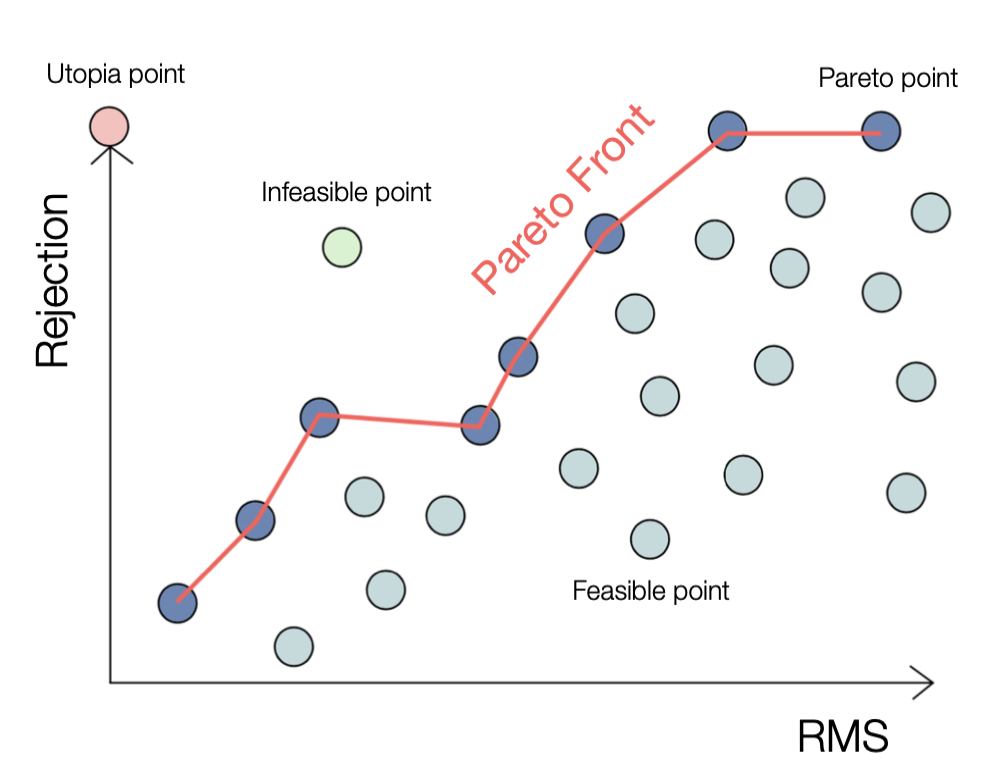}
     \caption{An example Pareto front plot. The purple points represent feasible points which maximize simultaneously the RMS and rejection values. The blue points are subdominant to the pareto front points, while the pink point represents the optimal desired result \cite{10.1007/978-3-540-88908-3_14}.} \label{fig:pareto-ex}
   \end{minipage}\hfill
   \begin{minipage}{0.48\textwidth}
     \centering
     \includegraphics[height= 5.3cm, keepaspectratio, trim=2mm 0mm 4mm 12mm, clip]{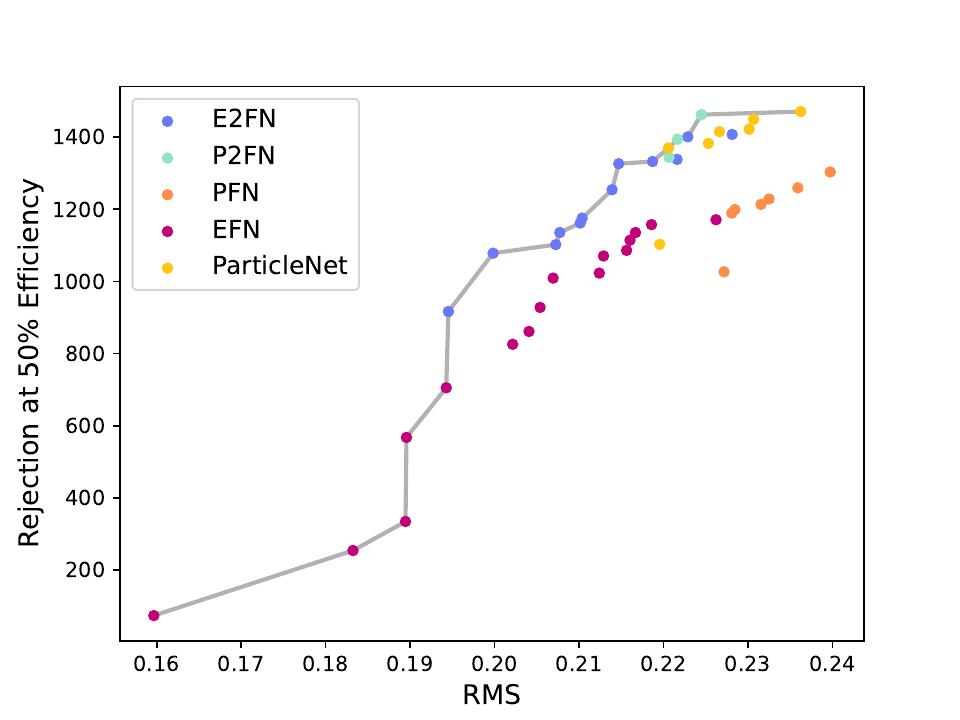}
     \caption{Rejection at 50\% efficiency vs RMS comparison between EFN, E2FN, PFN, P2FN, and ParticleNet.  E2FN and EFN dominate the pareto plot at low performance and high regularization, while ParticleNet and P2FN dominate at high performance and low regularization. }\label{fig:pareto}
   \end{minipage}
\end{figure}

The second study performed uses the RMS metric to compare the robustness of the NNs against their performances.  For it, the reclustering of all networks was set to 1.3 GeV and their performances were compared using Pareto Front plots.  This type of plots  are commonly used to visualize problems with multiple objectives, where neither objective is more important than the other.  In this manner, the pareto front represents the set of solutions which optimize all chosen objectives, see \Fig{fig:pareto-ex}.  In this study the objectives are maximizing the performance, thus increasing the rejection, and maximizing the robustness, hence decreasing the RMS.

As shown in \Fig{fig:pareto}, the Pareto front is dominated by the EFN at low performance and high robustness, followed by the E2FN at higher performance and lower robustness, then by the P2FN, and finally Particle Net at the highest performance and lowest robustness of all five studied networks.  By this metric the PFN is subdominant, meaning that at any given point in its training, there is another NN which outperforms it both in terms of robustness and performance.  The trend in \Fig{fig:pareto} indicates the same trade-off between performance and robustness as seen in \Fig{fig:recluster}, such that the level at which a network is affected by non-perturbative systematics will be dependent on the chosen working point.  Furthermore, this plot clearly lays out the shift from a wide range of working points which  would result in an improved robustness for the EFN and E2FN to a lack thereof for ParticleNet, P2FN, and PFN.
That is, although better performing, ParticleNet, PFN and P2FN are constrained to working points which are notably more susceptible to non-perturbative systematics. 

While the RMS score provides interesting insight into a NN's robustness, the robustness metric in and of itself does not directly correlate to any relevant physical quantities.  In order to gain a better understanding of the effects QCD systematics could have on an analysis, \Sec{sec:toys} presents a toy study which analyzes how a NN's robustness might affect the significance of a measurement.

\subsection{Impact of Robustness on Searches: A Toy Study}\label{sec:toys}

To highlight the impact that robustness can have on the choice of NN architecture for a particular search, in this section we perform a simple toy study. We consider a search for a Z boson, and rank networks by their performance including systematic uncertainties. We see that this significantly modifies the conclusions regarding which networks will be most performant, and highlights the importance of being able to tune between performance and robustness.

For our simple toy study, we consider the performance for tagging a boosted Z boson, as above, in the simplified scenario where the  NN represents the entire analysis work flow.  Thus, the only cuts performed are based on the NN's signal score, where the NN cut is chosen to optimize the analysis for best discovery significance.  This toy study was performed twice, with and without theoretical systematics.  As shown in \Fig{subfig:sig-a}, when performed disregarding any theoretical systematics, the discovery significance trends followed closely those seen in the raw performance plot, \Fig{fig:performance}, where the best discovery significance was achieved by the ParticleNet, followed by P2FN, the E2FN, PFN, and EFN, in that order. The study was then repeated when taking theoretical systematics into account, where the background systematics were taken to be 
\begin{equation}\label{sys-bkgr}
    \sigma_B = \frac{1}{10}\left[ B (\mathrm{hadron})- B (\mathrm{parton})\right]\,,
\end{equation}
where B is the number of events tagged as background by the NN when given hadron or parton-level data.  A tenth of this difference was chosen as the uncertainty, as it yielded a $\sim 20\%$ uncertainty on the data, which was deemed to be in the right range for theoretical uncertainties.  This study showed that, when taking theoretical uncertainties into account, the measurement's significance when using ParticleNet and PFN drastically decreased.  However, the significance for the toy measurements performed with the E2FN and EFN were considerably less affected \Fig{subfig:sig-b}.  This shows that IRC safety can indeed help protect a measurement against non-perturbative uncertainties. While this study is clearly quite simple in scope, we believe that it highlights the importance of incorporating modeling uncertainties when deciding on the most performant networks for physics searches. It also highlights the need for architectures where the sensitivity to poorly modeled physics of jets can be smoothly controlled.

One other intriguing feature of our study was the
unexpectedly good performance of the P2FN network. This result is surprising considering the P2FN's architecture is not inherently IRC safe.  However, it seems to achieve a good balance of performance and resilience. While the reasons behind this unexpected result are still being studied, one thing is clear: there might be other non-IRC safe NNs that are well-behaved and can outperfom IRC safe ones. 

\begin{figure}
    \centering
    \subfloat[Discovery significance disregarding theoretical systematics.]{
        \includegraphics[width=.45\linewidth]{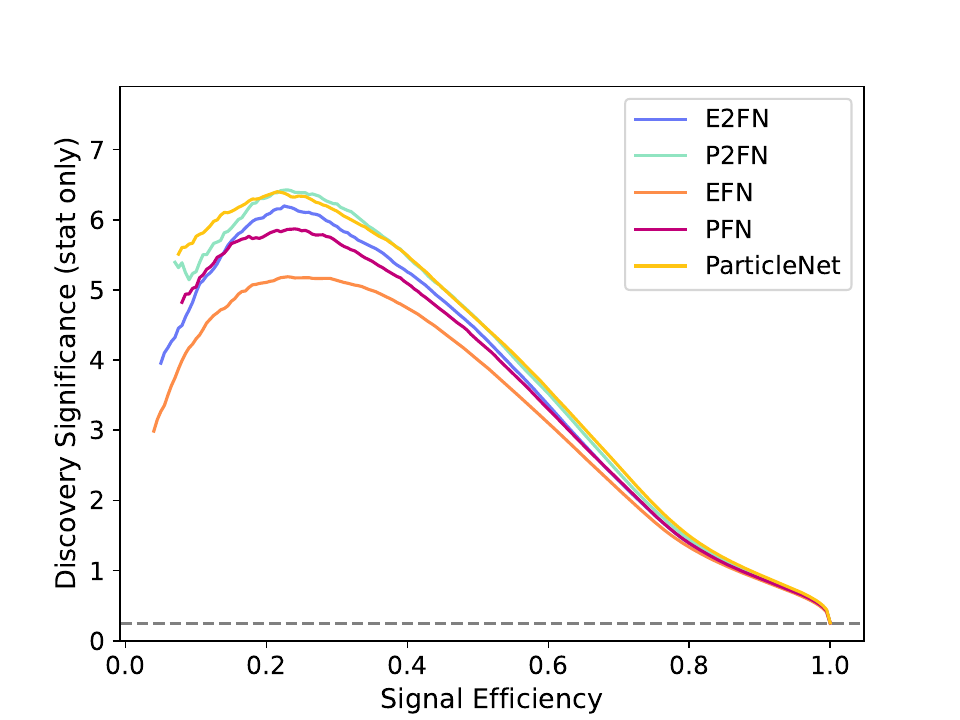}
        \label{subfig:sig-a}
    }\quad
    \subfloat[Discovery significance incorporating theoretical systematics.]{
        \includegraphics[width=.45\linewidth]{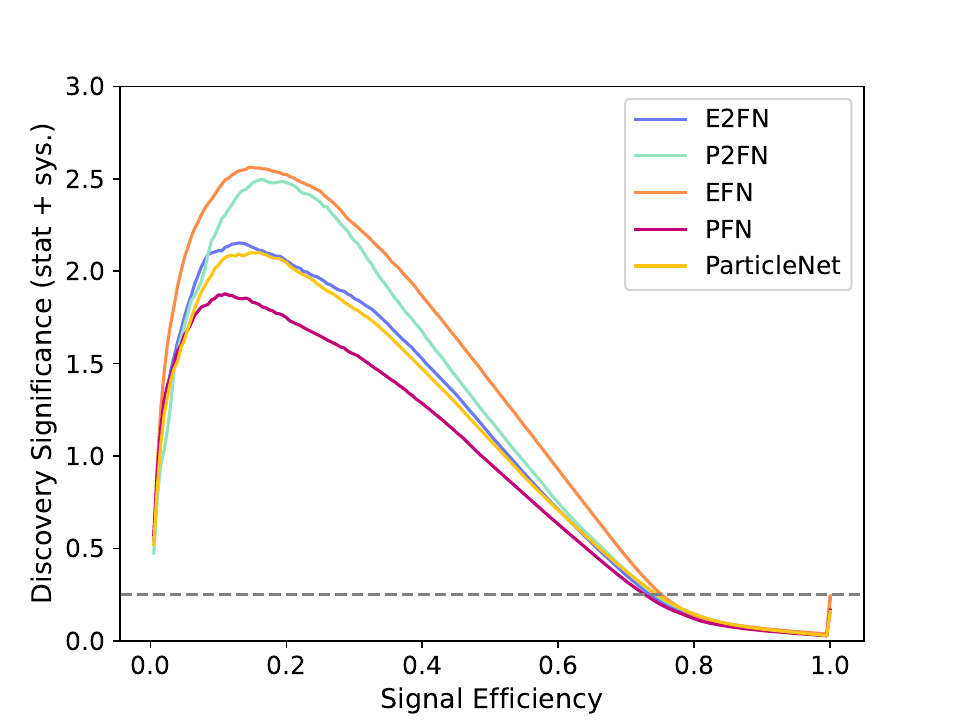}
        \label{subfig:sig-b}
    }
    \caption{Toy Study: tagging boosted Z bosons in QCD background. The NN was used as the entire work flow, where its score was cut on to optimize for discovery significance. \Fig{subfig:sig-a} shows the best discovery significance disregarding theoretical systematics. \Fig{subfig:sig-b} shows the best discovery significance taking into account background theoretical uncertainties \Eq{sys-bkgr}.}
    \label{fig:label}
\end{figure}

\section{Conclusions}\label{sec:conclusions}

The application of machine learning to jet substructure is, and will continue to be, one of the most exciting directions at the LHC. One important aspect of improving the understanding of machine learning applications in searches, is to ensure that uncertainties from the theoretical modeling of jets are under control. This is particularly important, since ML taggers use the detailed structure of jets well beyond what has been done previously. Along with advances in ML, there have also been tremendous advances in our ability to theoretically describe the internal structure of jets. Hence, it is important to be able to design architectures which maximally exploit these advances, instead of focusing on poorly understood non-perturbative effects.

In this paper we have taken several steps in this direction. First, we have introduced a new network architecture, named EnFN, which generalizes the traditional Energy Flow Networks (EFN) by directly incorporating  multi-particle correlations. We also introduced an analogous non-IRC safe version, the P2FN, which generalizes Particle Flow Networks (PFNs). These architectures are inspired by recent theoretical advances in the study of energy correlator observables. We studied the performance of these networks, focusing primarily on the case of the E2FN, which directly encodes two-point correlations in jets. The E2FN and, particularly, the P2FN showed promising performance in boosted Z discrimination, compared to the state-of-the-art ParticleNet.

A key aspect of this paper was to study the resilience of networks to hadronization effects in jets, introduce techniques to quantify these effects, and illustrate their impact on systematics for searches. By using Pythia events at both the parton level and hadron level, we were able quantify the sensitivity of the network to hadronization. We then introduced a convenient way of visualizing the performance vs. robustness of different architectures using so called ``Pareto plots". As expected, we found that the most performant networks were highly sensitive to non-perturbative corrections. The newly introduced E2FN networks have a much more minimal sensitivity to non-perturbative effects, which we further showed could be adjusted through the use of a pre-clustering stage.

Finally, we highlighted how uncertainties manifest in the discovery significance of a toy study, searching for a boosted Z boson. While most studies focus solely on tagging performance, taking into account modeling uncertainties, and ordering the networks by their achieved discovery significance drastically changes the picture, with less performant, but more resilient networks, achieving the best signficance.

Going forward, there are a number of directions which we think are important to pursue. We have highlighted how to construct networks that immediately focus on higher point correlations. It would be particularly interesting to study these in parton showers which incorporate higher point splitting functions to see if the improved theoretical description of the internal structure of jets enhances the network's ability to discriminate different signals. We also believe that it is important to explore in more detail the impact non-perturbative uncertainties have on different ML based searches. Having networks where the robustness vs. performance can be tuned in a controllable manner will be essential, as understanding and controlling uncertainties in ML based searches will allow us to maximize the discovery potential of many exciting searches at the LHC.

\acknowledgments
We thank Paul Tipton and Andrew Larkoski for useful discussions.
I.M. is supported by the the U.S. Department of Energy (DOE) Early Career Award DE-SC0025581, and the Sloan Foundation. A.G.C. and C.S. are supported by the DOE Office of High Energy Physics under grant number DE-SC0017660. The computations in this paper were performed on the Yale Grace computing cluster, supported by the facilities and staff at the Yale University Faculty of Sciences High Performance Computing Center.

\bibliography{EEC_ref.bib}

\providecommand{\href}[2]{#2}\begingroup\raggedright\begin{thebibliography}{10}

\bibitem{Butterworth:2008iy}
J.~M. Butterworth, A.~R. Davison, M.~Rubin, and G.~P. Salam, {\it {Jet substructure as a new Higgs search channel at the LHC}},  {\em Phys. Rev. Lett.} {\bf 100} (2008) 242001, [\href{http://arxiv.org/abs/0802.2470}{{\tt arXiv:0802.2470}}].

\bibitem{Larkoski:2017jix}
A.~J. Larkoski, I.~Moult, and B.~Nachman, {\it {Jet Substructure at the Large Hadron Collider: A Review of Recent Advances in Theory and Machine Learning}},  \href{http://arxiv.org/abs/1709.04464}{{\tt arXiv:1709.04464}}.

\bibitem{Asquith:2018igt}
R.~Kogler et~al., {\it {Jet Substructure at the Large Hadron Collider: Experimental Review}},  {\em Rev. Mod. Phys.} {\bf 91} (2019), no.~4 045003, [\href{http://arxiv.org/abs/1803.06991}{{\tt arXiv:1803.06991}}].

\bibitem{Marzani:2019hun}
S.~Marzani, G.~Soyez, and M.~Spannowsky, {\em {Looking inside jets: an introduction to jet substructure and boosted-object phenomenology}}, vol.~958.
\newblock Springer, 2019.

\bibitem{Thaler:2011gf}
J.~Thaler and K.~Van~Tilburg, {\it {Maximizing Boosted Top Identification by Minimizing N-subjettiness}},  {\em JHEP} {\bf 02} (2012) 093, [\href{http://arxiv.org/abs/1108.2701}{{\tt arXiv:1108.2701}}].

\bibitem{Thaler:2010tr}
J.~Thaler and K.~Van~Tilburg, {\it {Identifying Boosted Objects with N-subjettiness}},  {\em JHEP} {\bf 03} (2011) 015, [\href{http://arxiv.org/abs/1011.2268}{{\tt arXiv:1011.2268}}].

\bibitem{Larkoski:2014gra}
A.~J. Larkoski, I.~Moult, and D.~Neill, {\it {Power Counting to Better Jet Observables}},  {\em JHEP} {\bf 12} (2014) 009, [\href{http://arxiv.org/abs/1409.6298}{{\tt arXiv:1409.6298}}].

\bibitem{Moult:2016cvt}
I.~Moult, L.~Necib, and J.~Thaler, {\it {New Angles on Energy Correlation Functions}},  {\em JHEP} {\bf 12} (2016) 153, [\href{http://arxiv.org/abs/1609.07483}{{\tt arXiv:1609.07483}}].

\bibitem{Larkoski:2014zma}
A.~J. Larkoski, I.~Moult, and D.~Neill, {\it {Building a Better Boosted Top Tagger}},  {\em Phys. Rev. D} {\bf 91} (2015), no.~3 034035, [\href{http://arxiv.org/abs/1411.0665}{{\tt arXiv:1411.0665}}].

\bibitem{Larkoski:2013eya}
A.~J. Larkoski, G.~P. Salam, and J.~Thaler, {\it {Energy Correlation Functions for Jet Substructure}},  {\em JHEP} {\bf 06} (2013) 108, [\href{http://arxiv.org/abs/1305.0007}{{\tt arXiv:1305.0007}}].

\bibitem{ATLAS:2023nwp}
{\bf ATLAS} Collaboration, {\it {Constituent-Based Quark Gluon Tagging using Transformers with the ATLAS detector}}, .

\bibitem{ATLAS:2023zcb}
{\bf ATLAS} Collaboration, {\it {Constituent-Based $W$-boson Tagging with the ATLAS Detector}}, .

\bibitem{ATLAS:2023krw}
{\bf ATLAS} Collaboration, {\it {Tagging boosted $W$ bosons applying machine learning to the Lund Jet Plane}}, .

\bibitem{Duperrin:2023elp}
{\bf ATLAS} Collaboration, A.~Duperrin, {\it {Flavour tagging with graph neural networks with the ATLAS detector}},  in {\em {30th International Workshop on Deep-Inelastic Scattering and Related Subjects}}, 6, 2023.
\newblock \href{http://arxiv.org/abs/2306.04415}{{\tt arXiv:2306.04415}}.

\bibitem{ATLAS:2022qby}
{\bf ATLAS} Collaboration, {\it {Constituent-Based Top-Quark Tagging with the ATLAS Detector}}, .

\bibitem{CMS:2020poo}
{\bf CMS} Collaboration, A.~M. Sirunyan et~al., {\it {Identification of heavy, energetic, hadronically decaying particles using machine-learning techniques}},  {\em JINST} {\bf 15} (2020), no.~06 P06005, [\href{http://arxiv.org/abs/2004.08262}{{\tt arXiv:2004.08262}}].

\bibitem{ATLAS:2024doi}
{\bf ATLAS} Collaboration, {\it {Measurements of W H and Z H Higgs production with decays into bottom quarks and direct constraints on the charm Yukawa coupling with 13 TeV collisions in the ATLAS detector}}, .

\bibitem{ATLAS:2019lwq}
{\bf ATLAS} Collaboration, G.~Aad et~al., {\it {Identification of boosted Higgs bosons decaying into $b$-quark pairs with the ATLAS detector at 13 $\text {TeV}$}},  {\em Eur. Phys. J. C} {\bf 79} (2019), no.~10 836, [\href{http://arxiv.org/abs/1906.11005}{{\tt arXiv:1906.11005}}].

\bibitem{ATLAS:2024yzu}
{\bf ATLAS} Collaboration, G.~Aad et~al., {\it {Measurements of WH and ZH production with Higgs boson decays into bottom quarks and direct constraints on the charm Yukawa coupling in 13 TeV pp collisions with the ATLAS detector}},  {\em JHEP} {\bf 04} (2025) 075, [\href{http://arxiv.org/abs/2410.19611}{{\tt arXiv:2410.19611}}].

\bibitem{ATLAS:2023jdk}
{\bf ATLAS} Collaboration, G.~Aad et~al., {\it {Study of High-Transverse-Momentum Higgs Boson Production in Association with a Vector Boson in the qqbb Final State with the ATLAS Detector}},  {\em Phys. Rev. Lett.} {\bf 132} (2024), no.~13 131802, [\href{http://arxiv.org/abs/2312.07605}{{\tt arXiv:2312.07605}}].

\bibitem{ATLAS:2020jwz}
{\bf ATLAS} Collaboration, G.~Aad et~al., {\it {Measurement of the associated production of a Higgs boson decaying into $b$-quarks with a vector boson at high transverse momentum in $pp$ collisions at $\sqrt{s} = 13$ TeV with the ATLAS detector}},  {\em Phys. Lett. B} {\bf 816} (2021) 136204, [\href{http://arxiv.org/abs/2008.02508}{{\tt arXiv:2008.02508}}].

\bibitem{CMS:2020zge}
{\bf CMS} Collaboration, A.~M. Sirunyan et~al., {\it {Inclusive search for highly boosted Higgs bosons decaying to bottom quark-antiquark pairs in proton-proton collisions at $\sqrt{s} =$ 13 TeV}},  {\em JHEP} {\bf 12} (2020) 085, [\href{http://arxiv.org/abs/2006.13251}{{\tt arXiv:2006.13251}}].

\bibitem{ATLAS:2021nsx}
{\bf ATLAS} Collaboration, {\it {Study of Higgs-boson production with large transverse momentum using the $H\rightarrow b\bar{b}$ decay with the ATLAS detector}}, .

\bibitem{CMS:2022psv}
{\bf CMS} Collaboration, A.~Tumasyan et~al., {\it {Search for Higgs Boson Decay to a Charm Quark-Antiquark Pair in Proton-Proton Collisions at s=13{\,}{\,}TeV}},  {\em Phys. Rev. Lett.} {\bf 131} (2023), no.~6 061801, [\href{http://arxiv.org/abs/2205.05550}{{\tt arXiv:2205.05550}}].

\bibitem{CMS:2022gjd}
{\bf CMS} Collaboration, A.~Tumasyan et~al., {\it {Search for Nonresonant Pair Production of Highly Energetic Higgs Bosons Decaying to Bottom Quarks}},  {\em Phys. Rev. Lett.} {\bf 131} (2023), no.~4 041803, [\href{http://arxiv.org/abs/2205.06667}{{\tt arXiv:2205.06667}}].

\bibitem{CMS:2024fkb}
{\bf CMS} Collaboration, A.~Hayrapetyan et~al., {\it {Search for Higgs boson pair production with one associated vector boson in proton-proton collisions at $ \sqrt{s} $ = 13 TeV}},  {\em JHEP} {\bf 10} (2024) 061, [\href{http://arxiv.org/abs/2404.08462}{{\tt arXiv:2404.08462}}].

\bibitem{Kats:2024eaq}
Y.~Kats and E.~Ofir, {\it {From strange-quark tagging to fragmentation tagging with machine learning}},  {\em Phys. Rev. D} {\bf 111} (2025), no.~3 034003, [\href{http://arxiv.org/abs/2408.12377}{{\tt arXiv:2408.12377}}].

\bibitem{Blekman:2024wyf}
F.~Blekman, F.~Canelli, A.~De~Moor, K.~Gautam, A.~Ilg, A.~Macchiolo, and E.~Ploerer, {\it {Tagging more quark jet flavours at FCC-ee at 91 GeV with a transformer-based neural network}},  {\em Eur. Phys. J. C} {\bf 85} (2025), no.~2 165, [\href{http://arxiv.org/abs/2406.08590}{{\tt arXiv:2406.08590}}].

\bibitem{Nakai:2020kuu}
Y.~Nakai, D.~Shih, and S.~Thomas, {\it {Strange Jet Tagging}},  \href{http://arxiv.org/abs/2003.09517}{{\tt arXiv:2003.09517}}.

\bibitem{Andreassen:2019cjw}
A.~Andreassen, P.~T. Komiske, E.~M. Metodiev, B.~Nachman, and J.~Thaler, {\it {OmniFold: A Method to Simultaneously Unfold All Observables}},  {\em Phys. Rev. Lett.} {\bf 124} (2020), no.~18 182001, [\href{http://arxiv.org/abs/1911.09107}{{\tt arXiv:1911.09107}}].

\bibitem{H1:2023fzk}
{\bf H1} Collaboration, V.~Andreev et~al., {\it {Unbinned deep learning jet substructure measurement in high Q2ep collisions at HERA}},  {\em Phys. Lett. B} {\bf 844} (2023) 138101, [\href{http://arxiv.org/abs/2303.13620}{{\tt arXiv:2303.13620}}].

\bibitem{ATLAS:2025qtv}
{ATLAS Collaboration}, {\it {Measurement of jet track functions in pp collisions at s=13 TeV with the ATLAS detector}},  {\em Phys. Lett. B} {\bf 868} (2025) 139680, [\href{http://arxiv.org/abs/2502.02062}{{\tt arXiv:2502.02062}}].

\bibitem{Song:2023sxb}
{\bf STAR} Collaboration, Y.~Song, {\it {Measurement of CollinearDrop jet mass and its correlation with SoftDrop groomed jet substructure observables in $\sqrt{s}=200$ GeV $pp$ collisions by STAR}},  \href{http://arxiv.org/abs/2307.07718}{{\tt arXiv:2307.07718}}.

\bibitem{Schwartz:2021ftp}
M.~D. Schwartz, {\it {Modern Machine Learning and Particle Physics}},  \href{http://arxiv.org/abs/2103.12226}{{\tt arXiv:2103.12226}}.

\bibitem{Karagiorgi:2022qnh}
G.~Karagiorgi, G.~Kasieczka, S.~Kravitz, B.~Nachman, and D.~Shih, {\it {Machine learning in the search for new fundamental physics}},  {\em Nature Rev. Phys.} {\bf 4} (2022), no.~6 399--412.

\bibitem{Dasgupta:2020fwr}
M.~Dasgupta, F.~A. Dreyer, K.~Hamilton, P.~F. Monni, G.~P. Salam, and G.~Soyez, {\it {Parton showers beyond leading logarithmic accuracy}},  {\em Phys. Rev. Lett.} {\bf 125} (2020), no.~5 052002, [\href{http://arxiv.org/abs/2002.11114}{{\tt arXiv:2002.11114}}].

\bibitem{Hoche:2024dee}
S.~H\"oche, F.~Krauss, and D.~Reichelt, {\it {The Alaric parton shower for hadron colliders}},  \href{http://arxiv.org/abs/2404.14360}{{\tt arXiv:2404.14360}}.

\bibitem{Catani:1998nv}
S.~Catani and M.~Grazzini, {\it {Collinear factorization and splitting functions for next-to-next-to-leading order QCD calculations}},  {\em Phys. Lett. B} {\bf 446} (1999) 143--152, [\href{http://arxiv.org/abs/hep-ph/9810389}{{\tt hep-ph/9810389}}].

\bibitem{Campbell:1997hg}
J.~M. Campbell and E.~W.~N. Glover, {\it {Double unresolved approximations to multiparton scattering amplitudes}},  {\em Nucl. Phys. B} {\bf 527} (1998) 264--288, [\href{http://arxiv.org/abs/hep-ph/9710255}{{\tt hep-ph/9710255}}].

\bibitem{Hoche:2017iem}
S.~H\"oche and S.~Prestel, {\it {Triple collinear emissions in parton showers}},  {\em Phys. Rev. D} {\bf 96} (2017), no.~7 074017, [\href{http://arxiv.org/abs/1705.00742}{{\tt arXiv:1705.00742}}].

\bibitem{Chen:2019bpb}
H.~Chen, M.-X. Luo, I.~Moult, T.-Z. Yang, X.~Zhang, and H.~X. Zhu, {\it {Three point energy correlators in the collinear limit: symmetries, dualities and analytic results}},  {\em JHEP} {\bf 08} (2020), no.~08 028, [\href{http://arxiv.org/abs/1912.11050}{{\tt arXiv:1912.11050}}].

\bibitem{Komiske:2022enw}
P.~T. Komiske, I.~Moult, J.~Thaler, and H.~X. Zhu, {\it {Analyzing N-point Energy Correlators Inside Jets with CMS Open Data}},  \href{http://arxiv.org/abs/2201.07800}{{\tt arXiv:2201.07800}}.

\bibitem{Chen:2022swd}
H.~Chen, I.~Moult, J.~Thaler, and H.~X. Zhu, {\it {Non-Gaussianities in Collider Energy Flux}},  \href{http://arxiv.org/abs/2205.02857}{{\tt arXiv:2205.02857}}.

\bibitem{Chicherin:2024ifn}
D.~Chicherin, I.~Moult, E.~Sokatchev, K.~Yan, and Y.~Zhu, {\it {The Collinear Limit of the Four-Point Energy Correlator in $\mathcal{N} = 4$ Super Yang-Mills Theory}},  \href{http://arxiv.org/abs/2401.06463}{{\tt arXiv:2401.06463}}.

\bibitem{Proceedings:2018jsb}
{\em {Les Houches 2017: Physics at TeV Colliders Standard Model Working Group Report}}, 3, 2018.

\bibitem{Butter:2022xyj}
A.~Butter, B.~M. Dillon, T.~Plehn, and L.~Vogel, {\it {Performance versus resilience in modern quark-gluon tagging}},  {\em SciPost Phys. Core} {\bf 6} (2023) 085, [\href{http://arxiv.org/abs/2212.10493}{{\tt arXiv:2212.10493}}].

\bibitem{ATLAS:2024rua}
{\bf ATLAS} Collaboration, G.~Aad et~al., {\it {Accuracy versus precision in boosted top tagging with the ATLAS detector}},  {\em JINST} {\bf 19} (2024), no.~08 P08018, [\href{http://arxiv.org/abs/2407.20127}{{\tt arXiv:2407.20127}}].

\bibitem{Moult:2025nhu}
I.~Moult and H.~X. Zhu, {\it {Energy Correlators: A Journey From Theory to Experiment}},  \href{http://arxiv.org/abs/2506.09119}{{\tt arXiv:2506.09119}}.

\bibitem{Komiske:2018cqr}
P.~T. Komiske, E.~M. Metodiev, and J.~Thaler, {\it {Energy Flow Networks: Deep Sets for Particle Jets}},  {\em JHEP} {\bf 01} (2019) 121, [\href{http://arxiv.org/abs/1810.05165}{{\tt arXiv:1810.05165}}].

\bibitem{Chen:2020vvp}
H.~Chen, I.~Moult, X.~Zhang, and H.~X. Zhu, {\it {Rethinking jets with energy correlators: Tracks, resummation, and analytic continuation}},  {\em Phys. Rev. D} {\bf 102} (2020), no.~5 054012, [\href{http://arxiv.org/abs/2004.11381}{{\tt arXiv:2004.11381}}].

\bibitem{Qu:2019gqs}
H.~Qu and L.~Gouskos, {\it {ParticleNet: Jet Tagging via Particle Clouds}},  {\em Phys. Rev. D} {\bf 101} (2020), no.~5 056019, [\href{http://arxiv.org/abs/1902.08570}{{\tt arXiv:1902.08570}}].

\bibitem{10.1007/978-3-540-88908-3_14}
E.~Zitzler, J.~Knowles, and L.~Thiele, {\em Quality Assessment of Pareto Set Approximations}, pp.~373--404.
\newblock Springer-Verlag, Berlin, Heidelberg, 2008.

\bibitem{Gambhir:2025xim}
R.~Gambhir, M.~LeBlanc, and Y.~Zhou, {\it {The Pareto Frontier of Resilient Jet Tagging}},  in {\em {39th Annual Conference on Neural Information Processing Systems}: {Includes Machine Learning and the Physical Sciences (ML4PS)}}, 9, 2025.
\newblock \href{http://arxiv.org/abs/2509.19431}{{\tt arXiv:2509.19431}}.

\bibitem{Budhraja:2024xiq}
A.~Budhraja and W.~J. Waalewijn, {\it {FastEEC: Fast evaluation of N-point energy correlators}},  {\em Phys. Lett. B} {\bf 861} (2025) 139276, [\href{http://arxiv.org/abs/2406.08577}{{\tt arXiv:2406.08577}}].

\bibitem{Tkachov:1999py}
F.~V. Tkachov, {\it {A Theory of jet definition}},  {\em Int. J. Mod. Phys. A} {\bf 17} (2002) 2783--2884, [\href{http://arxiv.org/abs/hep-ph/9901444}{{\tt hep-ph/9901444}}].

\bibitem{Tkachov:1995kk}
F.~V. Tkachov, {\it {Measuring multi - jet structure of hadronic energy flow or What is a jet?}},  {\em Int. J. Mod. Phys. A} {\bf 12} (1997) 5411--5529, [\href{http://arxiv.org/abs/hep-ph/9601308}{{\tt hep-ph/9601308}}].

\bibitem{Sveshnikov:1995vi}
N.~Sveshnikov and F.~Tkachov, {\it {Jets and quantum field theory}},  {\em Phys. Lett. B} {\bf 382} (1996) 403--408, [\href{http://arxiv.org/abs/hep-ph/9512370}{{\tt hep-ph/9512370}}].

\bibitem{Sjostrand:2007gs}
T.~Sjostrand, S.~Mrenna, and P.~Z. Skands, {\it {A Brief Introduction to PYTHIA 8.1}},  {\em Comput. Phys. Commun.} {\bf 178} (2008) 852--867, [\href{http://arxiv.org/abs/0710.3820}{{\tt arXiv:0710.3820}}].

\bibitem{Sjostrand:2014zea}
T.~Sj\"ostrand, S.~Ask, J.~R. Christiansen, R.~Corke, N.~Desai, P.~Ilten, S.~Mrenna, S.~Prestel, C.~O. Rasmussen, and P.~Z. Skands, {\it {An introduction to PYTHIA 8.2}},  {\em Comput. Phys. Commun.} {\bf 191} (2015) 159--177, [\href{http://arxiv.org/abs/1410.3012}{{\tt arXiv:1410.3012}}].

\bibitem{Cacciari:2008gp}
M.~Cacciari, G.~P. Salam, and G.~Soyez, {\it {The anti-$k_t$ jet clustering algorithm}},  {\em JHEP} {\bf 04} (2008) 063, [\href{http://arxiv.org/abs/0802.1189}{{\tt arXiv:0802.1189}}].

\bibitem{Cacciari:2011ma}
M.~Cacciari, G.~P. Salam, and G.~Soyez, {\it {FastJet User Manual}},  {\em Eur. Phys. J. C} {\bf 72} (2012) 1896, [\href{http://arxiv.org/abs/1111.6097}{{\tt arXiv:1111.6097}}].

\bibitem{pytorch:2019neurips}
A.~Paszke, S.~Gross, F.~Massa, A.~Lerer, J.~Bradbury, G.~Chanan, T.~Killeen, Z.~Lin, N.~Gimelshein, L.~Antiga, A.~Desmaison, A.~Köpf, E.~Z. Yang, Z.~DeVito, M.~Raison, A.~Tejani, S.~Chilamkurthy, B.~Steiner, L.~Fang, J.~Bai, and S.~Chintala, {\it Pytorch: An imperative style, high-performance deep learning library.},  in {\em NeurIPS} (H.~M. Wallach, H.~Larochelle, A.~Beygelzimer, F.~d'Alché Buc, E.~B. Fox, and R.~Garnett, eds.), pp.~8024--8035, 2019.
\newblock \href{http://arxiv.org/abs/1912.01703}{{\tt arXiv:1912.01703}}.

\bibitem{Kasieczka:2020nyd}
G.~Kasieczka, S.~Marzani, G.~Soyez, and G.~Stagnitto, {\it {Towards Machine Learning Analytics for Jet Substructure}},  {\em JHEP} {\bf 09} (2020) 195, [\href{http://arxiv.org/abs/2007.04319}{{\tt arXiv:2007.04319}}].

\end{thebibliography}\endgroup
\bibliographystyle{JHEP}

\end{document}